\documentclass[journal,onecolumn]{IEEEtran}

\usepackage{amsmath}
\usepackage{amsfonts,amssymb,amsthm}
\usepackage{mathrsfs}
\usepackage{array}
\usepackage[caption=false,font=normalsize,labelfont=sf,textfont=sf]{subfig}
\usepackage{textcomp}
\usepackage{stfloats}
\usepackage{url}
\usepackage{verbatim}
\usepackage{graphicx}
\usepackage{subfig}
\usepackage{adjustbox}
\usepackage{cite}
\usepackage{booktabs}
\usepackage{multicol}
\usepackage{multirow}
\usepackage{empheq}
\usepackage{xcolor}
\usepackage{bm}
\usepackage{float}
\usepackage{diagbox} 
\usepackage{tikz}
\usetikzlibrary{arrows.meta,positioning}
\usepackage{makecell}
\usepackage{setspace}
\usepackage{needspace}
\usepackage{enumerate}

\usepackage[ruled,vlined,linesnumbered]{algorithm2e}

\SetAlCapNameFnt{\bfseries}
\SetAlCapFnt{}
\newtheorem{theorem}{Theorem}[section]
\newtheorem{lemma}{Lemma}[section]
\newtheorem{corollary}{Corollary}[section]
\newtheorem{proposition}{Proposition}[section]
\newtheorem{define}{Definition}[section]

\newtheorem{example}{Example}[section]
\newtheorem{property}{Property}[section]

\newtheorem{construction}{System}

\def \Acal{{\mathcal A}}
\def \Bcal{{\mathcal B}}
\def \Ccal{{\mathcal C}}
\def \Dcal{{\mathcal D}}

\def \Rcal{{\mathcal R}}
\def \Ccal{{\mathcal C}}

\def \Ncal{{\mathcal N}}

\def \CScal {{\mathcal{CS}}}

\newcommand{\chain}[1]{\mbox{{\sf chain}$(#1)$}}

\usepackage{fullpage}
\begin{document}

	\title{Toward the Minimal Set of Information Inequalities in Regenerating Codes: An Algebraic Approach}
    \author{
  Rui-Juan Jing$^*$, Laigang Guo$^\ddagger$, Yan-Feng Xie$^\dagger$$^\S$, Chun-Ming Yuan$^\dagger$$^\S$
  \thanks{
    Corresponding author:  Laigang Guo (Email: lgguo@bnu.edu.cn).
    This work was partially supported by the National Natural Science Foundation of China under Grant Nos. 12101267, 12271516 and 12301647.
  }\\
  \small 
  $^*$\textit{School of Mathematical Sciences, Jiangsu University}, Zhenjiang, Jiangsu, China  \\
  $^\ddagger$\textit{Laboratory of Mathematics and Complex Systems (Ministry of Education),}\\ 
  \textit{School of Mathematical Sciences, Beijing Normal University}, Beijing, China \\
  $^\dagger$\textit{State Key Laboratory of Mathematical Sciences, Academy of Mathematics and Systems Science,}\\
  \textit{Chinese Academy of Sciences}, Beijing, China \\
  $^\S$\textit{School of Mathematical Sciences, University of Chinese Academy of Sciences}, Beijing, China\\
  \textit{Emails}: rjing@ujs.edu.cn, lgguo@bnu.edu.cn, cmyuan@mmrc.iss.ac.cn, xieyanfeng@amss.ac.cn
}
	\maketitle

\begin{abstract}
Information-theoretic outer bounds for exact-repair regenerating codes can be formulated as linear programs, but the associated Shannon-type systems contain many implicit equalities and redundant inequalities. We directly specify a reduced system for the $(n,n-1,n-1)$ exact-repair model using the closure structure induced by its repair dependencies. Nonempty closed sets index the reduced joint-entropy coordinates, and inseparable edges determine the defining inequalities. We prove equivalence with the initial system by characterizing its implicit equalities and establishing sufficient redundancy criteria. We also give an algorithm that constructs the directly defined system for any $n$, count its entropy coordinates, and derive an explicitly computable upper bound on its number of inequalities. For every $n\ge2$, we prove that the reduced system has no implicit equalities and that each distinct defining inequality is nonredundant and defines a facet in the nonempty closed-set coordinate space. Exact enumerations for $n\in\{2,3,4,5\}$ verify the reported system sizes.
\end{abstract}


\section{Introduction}
\label{sec:intro}
Determining the fundamental limits of information systems is a central problem in information theory. Converse arguments express the requirements of a system as relations among entropies and combine them with information inequalities. Yeung's linear programming (LP) framework~\cite{Yeung1997} makes this approach computational: joint entropies are the LP variables, elemental inequalities supply generic constraints, and the model supplies additional equalities.

Regenerating codes provide an important setting for this approach. An $(n,k,d)$ code stores a file across $n$ nodes so that any $k$ nodes recover it, while a failed node is repaired by downloading data from $d$ surviving nodes~\cite{DGWWR2010}. Functional repair requires the replacement data to preserve these reconstruction and repair requirements; exact repair requires the replacement node to reproduce the data originally stored at the failed node. The storage--repair-bandwidth tradeoff under exact repair has been studied through code constructions and converse bounds~\cite{RSK2011,SRKR2012-1,SRKR2012-2,CJMRS2013,SPKVSK2016}.

Tian characterized the $(4,3,3)$ exact-repair rate region and established a gap from functional repair using a computer-aided converse based on information inequalities~\cite{Tian2014}. The $(5,4,4)$ region was subsequently characterized by a computational outer bound achievable by canonical layered codes~\cite{Tian2015}. These results illustrate how LP methods, building on the framework underlying ITIP~\cite{YY1996}, can both discover outer bounds and produce explicit converse proofs. Symmetry and other structural reductions extend the reach of these computations~\cite{Tian2018,TPHZ2021,ZT2018}. Chen and Tian~\cite{CT2022} further reformulated large converse LPs as maximin problems, solving a sequence of relaxed LPs to obtain a tighter outer bound for the $(6,5,5)$ problem. Such methods select inequalities for a particular bound; they also motivate studying how the entire constraint system can be simplified before an objective is chosen.

The study of information inequalities has developed corresponding computational and algebraic methods. Following ITIP, tools including Xitip and oXitip~\cite{PD2006,RDGPPYY2020}, minitip~\cite{Csirmaz2016}, psitip~\cite{Li2020}, and AITIP~\cite{HLTY2020} support their verification and manipulation. Guo \textit{et al.}~\cite{GYG2023} developed symbolic procedures for identifying implicit equalities and redundant inequalities; their subsequent work~\cite{GYG2025} uses Gaussian elimination to simplify information-inequality problems. The size of the underlying Shannon-type system remains a difficulty: $m$ random variables give $2^m-1$ joint-entropy coordinates and $m+\binom m2 2^{m-2}$ elemental inequalities.

We study the repair-constrained Shannon-type system for the $(n,n-1,n-1)$ model. Our analysis uses the algebraic structure of its entropy coordinates and linear constraints to describe a reduced system independently of a particular LP objective. Our specific contributions are as follows:
\begin{enumerate}
\item We characterize the elemental inequalities that become implicit equalities under the repair dependencies. A closure operation identifies equal entropy coordinates, and we give an exact formula for the number of closed sets.
\item We define the final system directly through the closed sets and their inseparable edges. The implicit-equality analysis and sufficient redundancy criteria prove that this system is equivalent to the initial system. They support the equivalence proof, rather than an iterative redundancy test used to define the final system. We separately prove that, for every $n\ge2$, the final system is full-dimensional and irredundant in its nonempty closed-set coordinates, with each inequality defining a facet.
\item We characterize inseparable edges and give algorithms for generating them and constructing the final system. Exact enumerations for $n\in\{2,3,4,5\}$ verify the reported coordinate and inequality counts.
\item We derive an exact local cover count from an off-diagonal complement digraph and use its row and column degrees to obtain a computable upper bound on the number of defining inequalities. We evaluate this bound for $n\in\{2,3,4,5\}$.
\end{enumerate}

Section~\ref{sec:mainidea} specifies the model and the initial system. Section~\ref{sec:contribution} defines the reduced system and proves equivalence, general irredundancy, and full dimensionality. Section~\ref{sec:edge} constructs its inseparable edges, and Section~\ref{sec:countbound} gives the counting bound. We first fix the notation and information inequalities used below.

All random variables are discrete, and their entropies are assumed finite. For $m$ random variables $X_1,\ldots,X_m$, let $[m]=\{1,\ldots,m\}$ and let $X_K=\{X_i:i\in K\}$ denote the indexed collection of random variables with indices in $K$. The notation $H(X_K)$ denotes their joint entropy. We use $H(X_\emptyset)=0$ and logarithms to base $2$. Conditional entropy and conditional mutual information are
\begin{align*}
&H(X\mid Y)=H(X,Y)-H(Y),\\
&I(X;Y\mid Z)=H(X,Z)+H(Y,Z)-H(X,Y,Z)-H(Z).
\end{align*}
Their nonnegativity gives the basic information inequalities. A linear inequality implied by these inequalities, together with specified linear equality constraints, is called a constrained Shannon-type inequality~\cite{Yeung1997,Yeung2008}.

\begin{theorem}[Elemental inequalities~\cite{Yeung1997}]
\label{element-ine}
Every basic information inequality is implied by
\[
\left\{\begin{aligned}
&H(X_i\mid X_{[m]\setminus\{i\}})\ge0,\quad i\in[m],\\
&I(X_i;X_j\mid X_K)\ge0,\quad i<j,\quad K\subseteq[m]\setminus\{i,j\}.
\end{aligned}\right.
\]
There are $u=m+\binom m2 2^{m-2}$ elemental inequalities for $m\ge2$.
\end{theorem}
For example, $H(X_1\mid X_3)=H(X_1\mid X_2,X_3)+I(X_1;X_2\mid X_3)$. Thus the elemental inequalities imply the nonnegativity of a conditional entropy with a smaller conditioning set.

Let $\mathbf h$ contain the $2^m-1$ nonempty joint entropies, and write the elemental inequalities as $\mathbf G\mathbf h\ge0$. Additional homogeneous equalities are written as $\mathbf Q\mathbf h=0$.
\begin{theorem}[LP characterization~\cite{Yeung1997}]
\label{LP-S}
The inequality $\mathbf b^\top\mathbf h\ge0$ is Shannon-type under $\mathbf Q\mathbf h=0$ if and only if
\[
\min\{\mathbf b^\top\mathbf h:\mathbf G\mathbf h\ge0,\ \mathbf Q\mathbf h=0\}=0.
\]
\end{theorem}
The zero vector is feasible, so validity makes the minimum zero. Conversely, a zero minimum implies nonnegativity on the feasible cone. The coordinate $H(X_\emptyset)=0$ is never counted as a variable.

\section{Problem reformulation}
	\label{sec:mainidea}

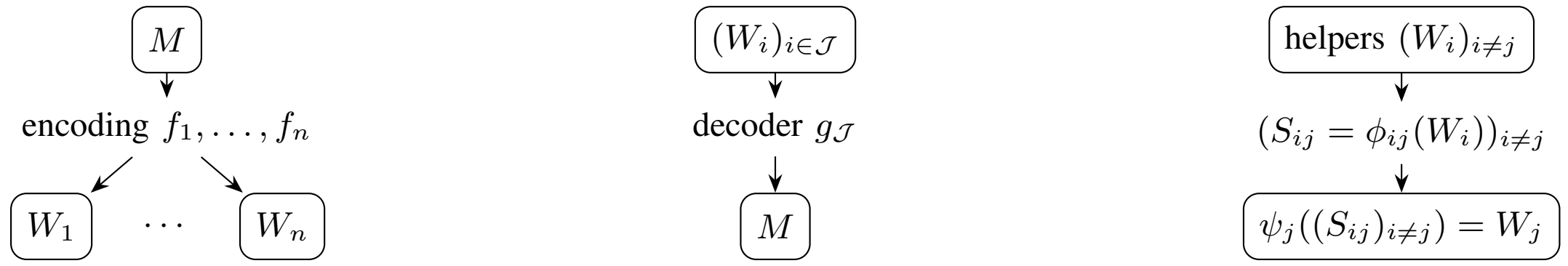
\begin{figure}[ht]

\centering
\begin{minipage}[t]{.31\linewidth}
\centering
\begin{tikzpicture}[>=Stealth,every node/.style={font=\small},box/.style={draw,rounded corners,minimum height=6mm,inner sep=4pt}]
\node[box] (m) at (0,1.7) {$M$};
\node (enc) at (0,.9) {encoding $f_1,\ldots,f_n$};
\node[box] (w1) at (-1.05,0) {$W_1$};
\node (dots) at (0,0) {$\cdots$};
\node[box] (wn) at (1.05,0) {$W_n$};
\draw[->] (m)--(enc);
\draw[->] (enc)--(w1);
\draw[->] (enc)--(wn);
\end{tikzpicture}
\par\small (a) Storage at $n$ nodes
\end{minipage}\hfill
\begin{minipage}[t]{.31\linewidth}
\centering
\begin{tikzpicture}[>=Stealth,every node/.style={font=\small},box/.style={draw,rounded corners,minimum height=6mm,inner sep=4pt}]
\node[box] (w) at (0,1.7) {$(W_i)_{i\in\mathcal J}$};
\node (dec) at (0,.9) {decoder $g_{\mathcal J}$};
\node[box] (m) at (0,0) {$M$};
\draw[->] (w)--(dec);
\draw[->] (dec)--(m);
\end{tikzpicture}
\par\small (b) Any $|\mathcal J|=n-1$ nodes
\end{minipage}\hfill
\begin{minipage}[t]{.33\linewidth}
\centering
\begin{tikzpicture}[>=Stealth,every node/.style={font=\small},box/.style={draw,rounded corners,minimum height=6mm,inner sep=4pt}]
\node[box] (w) at (0,1.7) {helpers $(W_i)_{i\ne j}$};
\node (s) at (0,.85) {$(S_{ij}=\phi_{ij}(W_i))_{i\ne j}$};
\node[box] (r) at (0,0) {$\psi_j((S_{ij})_{i\ne j})=W_j$};
\draw[->] (w)--(s);
\draw[->] (s)--(r);
\end{tikzpicture}
\par\small (c) Exact repair of node $j$
\end{minipage}
\caption{The $(n,n-1,n-1)$ exact-repair model: storage, reconstruction, and repair.}
\label{fig:erc-model}
\end{figure}

We consider the $(n,n-1,n-1)$ exact-repair model with $n\ge2$, following the functional description in~\cite{DGWWR2010,Tian2014}. A file $M$ is encoded into the contents $W_i=f_i(M)$ of $n$ storage nodes. Any $n-1$ nodes reconstruct the file: for every $\mathcal J\subseteq[n]$ with $|\mathcal J|=n-1$, a decoder satisfies $g_{\mathcal J}((W_i)_{i\in\mathcal J})=M$. To repair node $j$, every other node $i$ sends $S_{ij}=\phi_{ij}(W_i)$, and the replacement node applies a decoder $\psi_j$ satisfying $\psi_j((S_{ij})_{i\ne j})=W_j$. Figure~\ref{fig:erc-model} summarizes these operations.

The two repair operations give
\[
H((S_{ij})_{j\ne i}\mid W_i)=0,\qquad
H(W_j\mid(S_{ij})_{i\ne j})=0.
\]
These repair dependencies involve only the stored and repair variables $W_i$ and $S_{ij}$, whose total number is $n^2$. We study the Shannon-type system determined by these repair dependencies, without adding the file $M$ or imposing the encoding and file-reconstruction constraints in Fig.~\ref{fig:erc-model}(a) and (b). Every exact-repair code satisfies the repair dependencies, irrespective of its file size, storage limits, or bandwidth limits. Reducing this common subsystem therefore gives a representation that can be reused across converse problems with different additional constraints. Any additional constraints expressed in its entropy coordinates can subsequently be imposed on either equivalent representation. This separates the structural reduction from the choice of a particular converse LP.

Writing each zero conditional entropy as a difference of joint entropies gives the following repair equalities:
\begin{subequations}\label{eq:equations}
\begin{empheq}[left=\empheqlbrace]{align}
&H(W_i,(S_{ij})_{j\ne i})-H(W_i)=0,\quad i\in[n],\label{eqn:diag}\\
&H(W_j,(S_{ij})_{i\ne j})-H((S_{ij})_{i\ne j})=0,\quad j\in[n].\label{eqn:col}
\end{empheq}
\end{subequations}

The random-variable collection is
\[
\{W_i:1\le i\le n\}\ \cup\ \{S_{ij}:1\le i,j\le n,\ i\ne j\},
\]
so there are $N=n+n(n-1)=n^2$ random variables. Define $\mathcal N=\{1,\ldots,N\}$. For any subset of this collection, its joint entropy is an entropy coordinate. We relabel the random variables by
\[
X_{i+n(i-1)}=W_i,\qquad X_{j+n(i-1)}=S_{ij}\quad(i\ne j).
\]

The relabeling is displayed in the following $n\times n$ array:
		\begin{equation}\label{eq:index-array}
\left(\begin{matrix}
W_1&S_{12}&\cdots&S_{1n}\\
S_{21}&W_2&\cdots&S_{2n}\\
\vdots&\vdots&\vdots&\vdots\\
S_{n1}&S_{n2}&\cdots&W_n
\end{matrix}\right)
\quad\Longrightarrow\quad
\left(\begin{matrix}
X_1&X_2&\cdots&X_n\\
X_{n+1}&X_{n+2}&\cdots&X_{2n}\\
\vdots&\vdots&\vdots&\vdots\\
X_{n(n-1)+1}&X_{n(n-1)+2}&\cdots&X_{n^2}
\end{matrix}\right).
\end{equation}

	For $\mathcal A\subseteq\mathcal N$, let $X_{\mathcal A}=\{X_i:i\in\mathcal A\}$ denote the indexed collection of random variables selected by $\mathcal A$, and abbreviate their joint entropy $H(X_{\mathcal A})$ as $H_{\mathcal A}$.
	By Theorem \ref{element-ine}, we obtain the basic constraints (elemental inequalities):
	\begin{subequations}\label{eq:inequalities}
		\begin{empheq}[left=\empheqlbrace]{align}
&H(X_{\mathcal N})-H(X_{\mathcal N\setminus\{i\}})\ge0,\label{ineq:two}\\
&H(X_{\{i\}\cup\mathcal A})+H(X_{\{j\}\cup\mathcal A})-H(X_{\{i,j\}\cup\mathcal A})-H(X_{\mathcal A})\ge0.\label{ineq:four}
\end{empheq}
	\end{subequations}
	for $i,j\in\mathcal N$ with $i\ne j$ and $\mathcal A\subseteq\mathcal N\setminus\{i,j\}$.
	This elemental system has $2^N-1=2^{n^2}-1$ nonempty entropy coordinates and $N+\binom N2 2^{N-2}=O(2^{n^2}n^4)$ inequalities.
For the array~\eqref{eq:index-array}, let $\mathcal D_k$, $\mathcal R_k$, and $\mathcal C_k$ denote the index sets of its $k$th diagonal entry, its $k$th row excluding the diagonal, and its $k$th column excluding the diagonal, respectively:
\begin{align*}
&\mathcal D_k=\{n(k-1)+k\},\\
&\mathcal R_k=\{n(k-1)+1,\ldots,nk\}\setminus\mathcal D_k,\\
&\mathcal C_k=\{k,n+k,\ldots,n(n-1)+k\}\setminus\mathcal D_k.
\end{align*}
In particular, $\mathcal D_k$ is a singleton. By Theorem~\ref{element-ine} and the repair equalities~\eqref{eq:equations}, the initial Shannon-type system is as follows.

\begin{construction}
\label{const:initial}
\begin{subequations}\label{sys:initial}
\begin{empheq}[left=\empheqlbrace]{align}
&H_{\mathcal N}-H_{\mathcal N\setminus\{i\}}\ge0,\quad i\in\mathcal N,\label{ineq:twonew}\\
&H_{\{i\}\cup\mathcal A}+H_{\{j\}\cup\mathcal A}-H_{\{i,j\}\cup\mathcal A}-H_{\mathcal A}\ge0,
\quad i,j\in\mathcal N,\ i\ne j,\quad\mathcal A\subseteq\mathcal N\setminus\{i,j\},\label{ineq:fournew}\\
&H_{\mathcal D_k}=H_{\mathcal D_k\cup\mathcal R_k},\quad k\in[n],\label{eqn:diagnew}\\
&H_{\mathcal C_k}=H_{\mathcal D_k\cup\mathcal C_k},\quad k\in[n].\label{eqn:colnew}
\end{empheq}
\end{subequations}
\end{construction}

Here~\eqref{ineq:twonew} and~\eqref{ineq:fournew} are elemental inequalities, whereas~\eqref{eqn:diagnew} and~\eqref{eqn:colnew} are additional repair equalities. Imposing the latter on the elemental system forces many of its inequalities to become implicit equalities and makes others redundant. This makes a smaller equivalent representation possible. Our objective is to specify such a representation directly from the structure of System~\ref{const:initial} and prove that it defines the same feasible region.

\Needspace{9\baselineskip}
\section{Closure-based equivalent system}
\label{sec:contribution}
We first describe how the repair dependencies identify equal entropy coordinates and introduce the closure operation that records these identifications. This gives the equivalent intermediate System~\ref{thm:eliminatedsys}. We then use the closed sets and their inseparable edges to define \emph{System~\ref{const:final}} directly. The implicit-equality analysis and redundancy criteria establish equivalence with System~\ref{const:initial}. A separate argument then proves irredundancy and the absence of implicit equalities for every $n\ge2$. All entropy relations in this section are evaluated under System~\ref{const:initial} unless another system is specified.

\subsection{Implicit equalities}
	\label{sec:impliciteqs}

\begin{define}[Implicit equalities and redundant inequalities]\label{def:implicit-redundant}
An inequality $f(\mathbf h)\ge0$ in a feasible system is an \emph{implicit equality} if $f(\mathbf h)=0$ at every feasible point. It is \emph{redundant} if it is implied by the remaining constraints of the system.
\end{define}
Example~\ref{ex:impliciteqs} shows how the repair equalities force elemental inequalities to become implicit equalities.

	\begin{example}[Implicit equalities]
		\label{ex:impliciteqs}
		Let $n=3$, so that $N=9$ and $\mathcal N=\{1,\ldots,9\}$. 
		Starting with $\mathcal A_1=\{1,2\}$ and $\mathcal B_1=\{1\}$, define $\mathcal A_{i+1}=\mathcal A_i\cup\{i+3\}$ for $1\le i\le6$ and $\mathcal B_{i+1}=\mathcal B_i\cup\{i+2\}$ for $1\le i\le7$. Consider the following inequalities.

\begin{equation}\label{ineq:eqfirstpart}
\left\{\begin{aligned}
&H_{\Acal_i\cup\{3\}}+H_{\Acal_{i+1}}-H_{\Acal_{i+1}\cup\{3\}}-H_{\Acal_i}\ge0,\quad1\le i\le6,\\
&H_{\Ncal}-H_{\Acal_7}\ge0.
\end{aligned}\right.
\end{equation}

\begin{equation}\label{ineq:eqsecondpart}
\left\{\begin{aligned}
&H_{\Bcal_i\cup\{2\}}+H_{\Bcal_{i+1}}-H_{\Bcal_{i+1}\cup\{2\}}-H_{\Bcal_i}\ge0,\quad1\le i\le7,\\
&H_{\Ncal}-H_{\Bcal_8}\ge0.
\end{aligned}\right.
\end{equation}

Each left-hand side in the first line of~\eqref{ineq:eqfirstpart} is the difference
\[
\bigl(H_{\mathcal A_i\cup\{3\}}-H_{\mathcal A_i}\bigr)
-\bigl(H_{\mathcal A_{i+1}\cup\{3\}}-H_{\mathcal A_{i+1}}\bigr).
\]
Consequently, adjacent terms cancel when summed over $i$. Since $\mathcal A_1=\{1,2\}$, $\mathcal A_7=\mathcal N\setminus\{3\}$, and $\mathcal A_7\cup\{3\}=\mathcal N$, adding the last inequality gives
\begin{align*}
&\sum_{i=1}^{6}\bigl(H_{\mathcal A_i\cup\{3\}}+H_{\mathcal A_{i+1}}
-H_{\mathcal A_{i+1}\cup\{3\}}-H_{\mathcal A_i}\bigr)
+H_{\mathcal N}-H_{\mathcal A_7}\\
&\qquad=H_{\mathcal A_1\cup\{3\}}-H_{\mathcal A_1}
-H_{\mathcal A_7\cup\{3\}}+H_{\mathcal A_7}+H_{\mathcal N}-H_{\mathcal A_7}
=H_{1,2,3}-H_{1,2}.
\end{align*}
Likewise, $\mathcal B_1=\{1\}$ and $\mathcal B_8=\mathcal N\setminus\{2\}$ give
\begin{align*}
&\sum_{i=1}^{7}\bigl(H_{\mathcal B_i\cup\{2\}}+H_{\mathcal B_{i+1}}
-H_{\mathcal B_{i+1}\cup\{2\}}-H_{\mathcal B_i}\bigr)
+H_{\mathcal N}-H_{\mathcal B_8}\\
&\qquad=H_{\mathcal B_1\cup\{2\}}-H_{\mathcal B_1}
-H_{\mathcal B_8\cup\{2\}}+H_{\mathcal B_8}+H_{\mathcal N}-H_{\mathcal B_8}
=H_{1,2}-H_1.
\end{align*}
The sum of all $7+8=15$ nonnegative left-hand sides is $H_{1,2,3}-H_1=0$ by~\eqref{eqn:diagnew}, so every summand is zero. Thus one repair equality makes all fifteen displayed elemental inequalities implicit equalities, illustrating how many implicit equalities can arise from a single repair dependency. The next example examines the index patterns behind such equal entropy coordinates.

\end{example}

\begin{example}[Index patterns of implicit equalities]
		\label{ex:equivalence}
		For $n=3$, the row-1 repair equality~\eqref{eqn:diagnew} is $H_{\{1\}}=H_{\{1,2,3\}}$. Adjoining $2$ or $3$ to an index set that contains $1$ preserves its joint entropy, including when subsets of $\{4,7\}$ are present:
		\begin{equation*}
			\left\{
			\begin{aligned}
				&H_{\{1\}} = H_{\{1,2\}} = H_{\{1,3\}} = H_{\{1,2,3\}},\\
				&H_{\{1,4\}} = H_{\{1,2,4\}} = H_{\{1,3,4\}} = H_{\{1,2,3,4\}},\\
				&H_{\{1,7\}} = H_{\{1,2,7\}} = H_{\{1,3,7\}} = H_{\{1,2,3,7\}},\\
				&H_{\{1,4,7\}} = H_{\{1,2,4,7\}} = H_{\{1,3,4,7\}} = H_{\{1,2,3,4,7\}}.\\
			\end{aligned}
			\right.
		\end{equation*}

In each row of this display, adjoining any subset of $\{2,3\}$ to an index set containing $1$ leaves its entropy unchanged. This is the index pattern induced by the row-1 dependency. The column-1 equality~\eqref{eqn:colnew}, $H_{\{4,7\}}=H_{\{1,4,7\}}$, supplies a second pattern: $\{4,7\}$ determines $1$, which in turn determines $2$ and $3$. Hence
\[
H_{\{4,7\}}=H_{\{1,4,7\}}=H_{\{1,2,4,7\}}
=H_{\{1,3,4,7\}}=H_{\{1,2,3,4,7\}}.
\]
Both patterns express entropy-preserving enlargements of the index set. Their repeated application leads to a closed representative.

	\end{example}

The following definition formalizes these index patterns. Lemma~\ref{thm:closure} will show that replacing an index set by its closure preserves its entropy.

	\begin{define}[Closed Set and Closure]\label{def:closure}

Let $\mathcal A\subseteq\mathcal N$. We call $\mathcal A$ a \textbf{closed set} if, for every $k\in\{1,\ldots,n\}$,
\begin{enumerate}
\item $\mathcal D_k\subseteq\mathcal A$ implies $\mathcal R_k\subseteq\mathcal A$;
\item $\mathcal C_k\subseteq\mathcal A$ implies $\mathcal D_k\cup\mathcal R_k\subseteq\mathcal A$.
\end{enumerate}
The \textbf{closure} $\overline{\mathcal A}$ is the smallest closed set containing $\mathcal A$, with respect to inclusion. It exists and is unique: $\mathcal N$ is closed, and the intersection of any family of closed sets satisfies both defining conditions. Thus $\overline{\mathcal A}$ is the intersection of all closed supersets of $\mathcal A$. 

\end{define}
Let $\mathcal{CS}$ denote the collection of closed sets, including $\emptyset$ and $\mathcal N$. The nonempty members will index the reduced entropy coordinates, while $H_\emptyset=0$ is fixed.

\begin{figure}[H]
		\centering
		\subfloat[Row Criterion (R)        \label{fig:rowrule}]{    \includegraphics[width=0.4\linewidth]{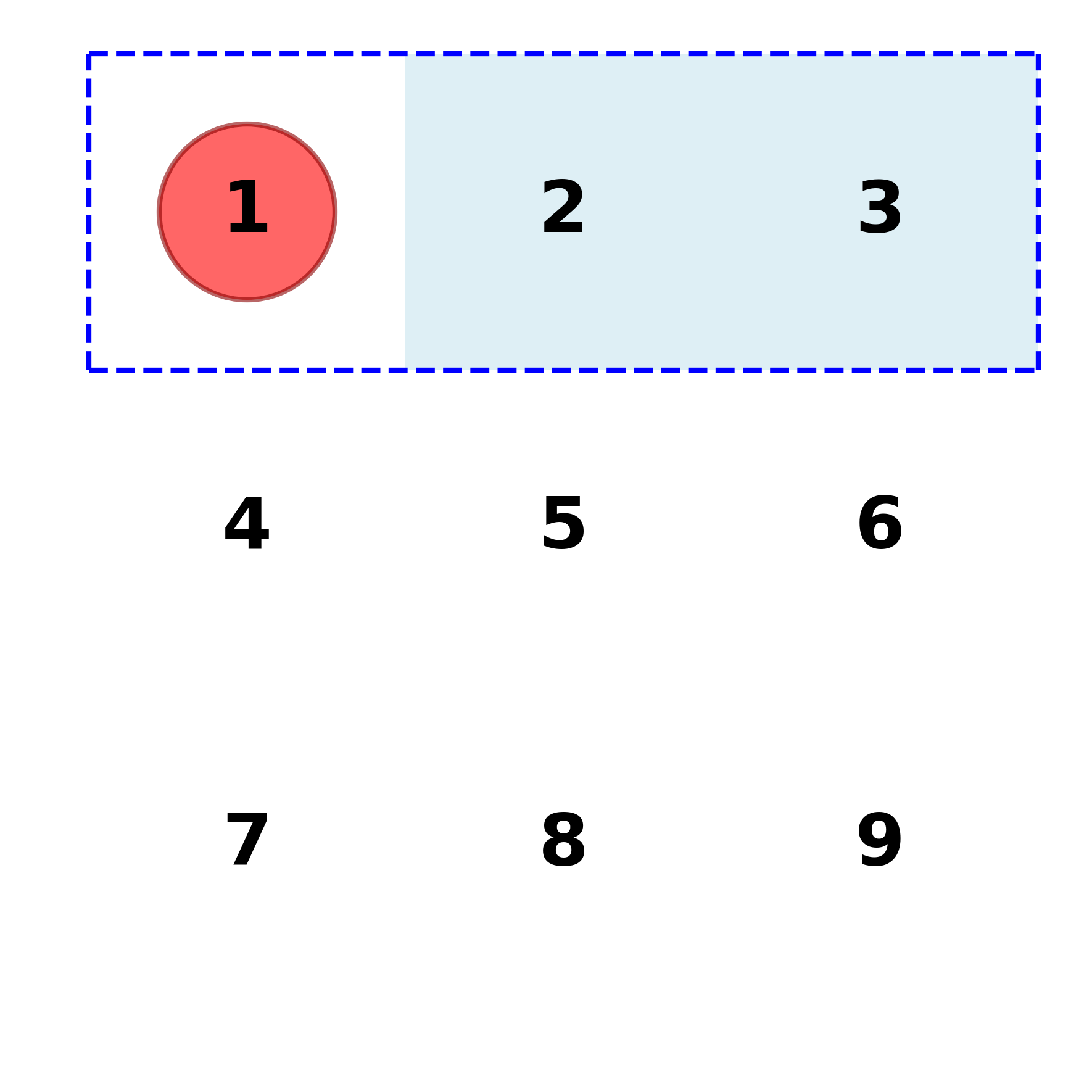} }
		\quad
		\subfloat[Column Criterion (C) \label{fig:columnrule}]{\includegraphics[width=0.4\linewidth]{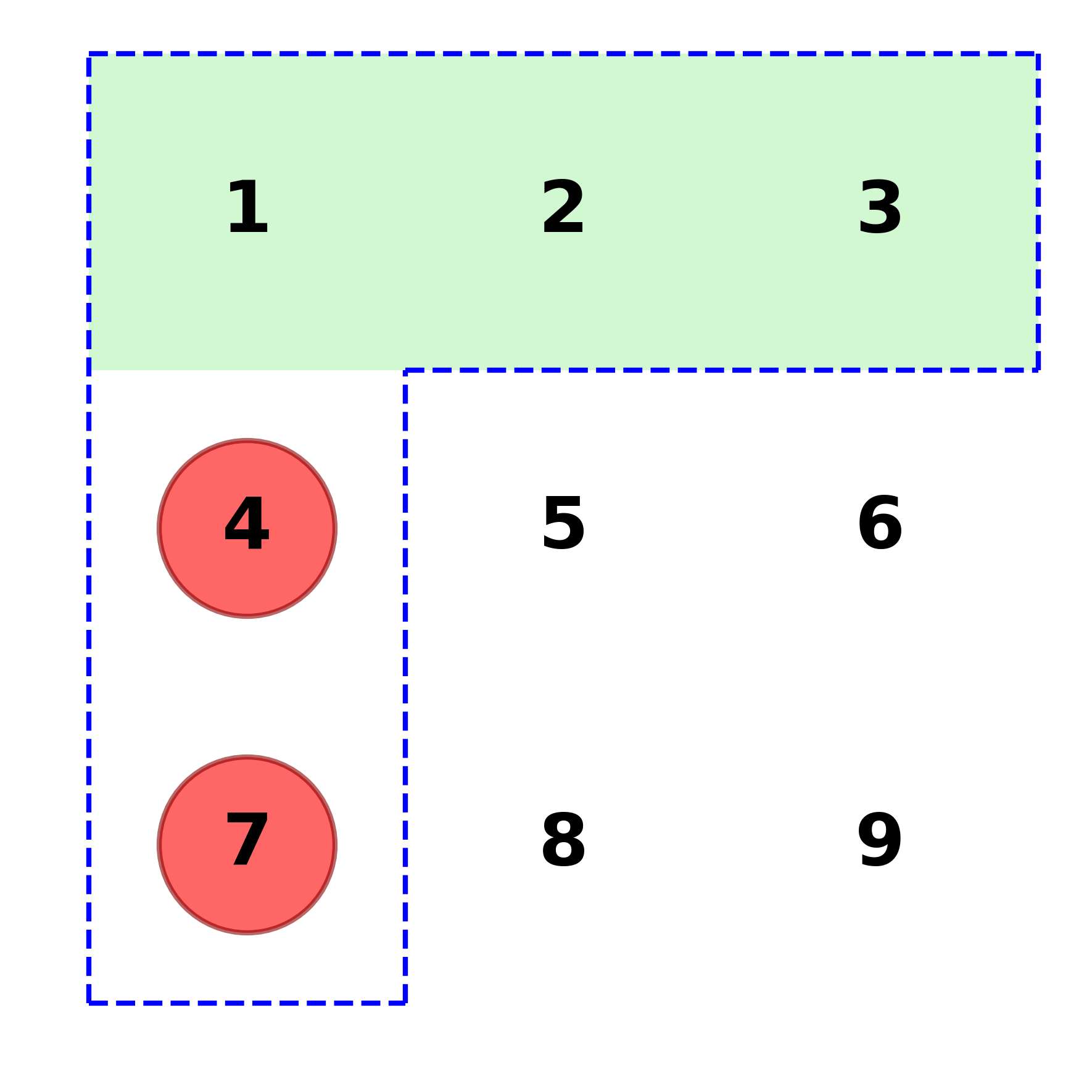}
		}

		\caption{Row and Column Criteria for Constructing Closed Sets}
		\label{fig:closesetrule}
	\end{figure}

Definition~\ref{def:closure} gives two rules for constructing closure, illustrated in Fig.~\ref{fig:closesetrule}:
	\begin{itemize}
		\item[(R)] if $\Dcal_k\subseteq\Acal$ for some $k$, then add $\Rcal_k$ into $\Acal$ (Fig.~\ref{fig:rowrule});
		\item[(C)] if $\Ccal_k\subseteq \Acal$ for some $k$, then add $\Dcal_k\cup\Rcal_k$ into $\Acal$ (Fig.~\ref{fig:columnrule}).
	\end{itemize}

Example~\ref{ex:equivalence} has closed representatives $\{1,2,3\}$, $\{1,2,3,4\}$, $\{1,2,3,7\}$, and $\{1,2,3,4,7\}$. Example~\ref{ex:constructionofclosure} illustrates the closure construction.
\begin{example}[Construction of Closure]\label{ex:constructionofclosure}
		Let $n=4$ and $\mathcal{A} = \{1, 7, 15\}$ be the initial set.
		Starting with the diagonal element $\mathcal{D}_1 = \{1\}$, we add the first row $\mathcal{R}_1 = \{2, 3, 4\}$ to the set.
		The newly included element $3$ completes $\mathcal C_3=\{3,7,15\}$, so the column criterion adds $\mathcal D_3\cup\mathcal R_3=\{9,10,11,12\}$.
		The process continues iteratively until closure is reached, yielding 
		$\overline{\mathcal{A}} = \{1, 2, 3, 4, 7, 9, 10, 11, 12, 15\}.$
		The complete construction is visualized in Fig.~\ref{fig:constructionofclosure}.
		\begin{figure}[H]

			\centering
			\includegraphics[width=.4\linewidth]{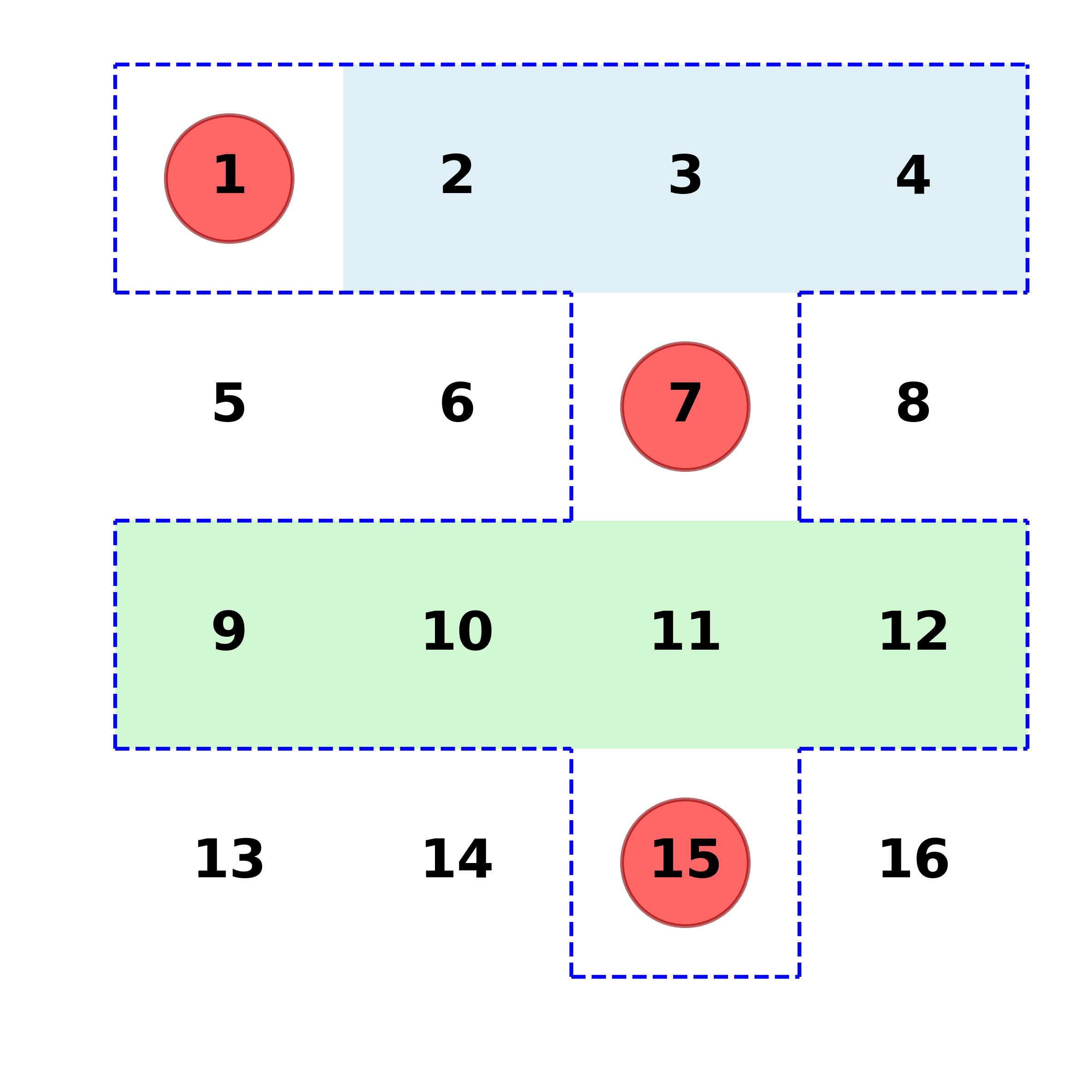} 
			\caption{Construct the closure for the index set $\Acal$.}
			\label{fig:constructionofclosure}
		\end{figure}

	\end{example}


Repeatedly applying the row and column criteria (R) and (C) until neither adds an index computes $\overline{\mathcal A}$. Each addition belongs to every closed superset of the initial set, and the process terminates because $\mathcal N$ is finite; the terminal set is therefore its smallest closed superset.

\Needspace{4\baselineskip}
We use the following properties of this closure operation.

\begin{property}[Closure Properties]
		\label{prop:closedset}
		Let $\mathcal A,\mathcal B\subseteq\mathcal N$ and $i,j\in\mathcal N$. The following properties hold:
		\begin{enumerate}
			\item \label{def:intersectionclosed}
			If $\mathcal{A}$ and $\mathcal{B}$ are closed, so is their intersection $\mathcal{A} \cap \mathcal{B}$;

			\item \label{def:closedsetinclusion}
			If $\mathcal{A} \subseteq \mathcal{B}$, then $\overline{\mathcal{A}} \subseteq \overline{\mathcal{B}}$;

			\item \label{def:elementinclusion}
			$\{i\} \cup \mathcal{A} \subseteq \{i\} \cup \overline{\mathcal{A}} \subseteq  \overline{\{i\} \cup \overline{\mathcal{A}}} = \overline{\{i\} \cup \mathcal{A}};
			$

			\item \label{thm:cap}
			If $\mathcal A$ is closed and $i,j\notin\mathcal A$ with $i\ne j$, then
			\begin{equation}
				\overline{\{i\} \cup \mathcal{A}} \cap \overline{\{j\} \cup \mathcal{A}} =
				\begin{cases}
					\overline{\{i\} \cup \mathcal{A}}, & \text{if } i \in \overline{\{j\} \cup \mathcal{A}}; \\[4pt]
					\overline{\{j\} \cup \mathcal{A}}, & \text{if } j \in \overline{\{i\} \cup \mathcal{A}}; \\[4pt]
					\mathcal{A}, & \text{otherwise}.
				\end{cases}
			\end{equation}
		\end{enumerate}
	\end{property}
\begin{proof}
The first three properties follow from the definition of closure. For the fourth, put $\mathcal B=\overline{\mathcal A\cup\{i\}}$ and $\mathcal C=\overline{\mathcal A\cup\{j\}}$. If $i\in\mathcal C$, monotonicity gives $\mathcal B\subseteq\mathcal C$; the case $j\in\mathcal B$ is symmetric.

It remains to consider $i\notin\mathcal C$ and $j\notin\mathcal B$. Suppose $(\mathcal B\cap\mathcal C)\setminus\mathcal A$ is nonempty. Any common new element differs from the seeds $i,j$ and is therefore added by completing its row in both closure constructions. Choose the first such common row $k$ in the construction of $\mathcal B$.

Neither construction can introduce row $k$ through its seed diagonal: that would put the seed in the other closure. Thus both complete row $k$ by the column criterion. Immediately before this step in the construction of $\mathcal B$, every element of $\mathcal C_k$ is already present; also $\mathcal C_k\subseteq\mathcal C$. Because $\mathcal A$ is closed and row $k$ is not already full in $\mathcal A$, there exists $x\in\mathcal C_k\setminus\mathcal A$. This $x$ is neither seed, so its row must have been completed in both constructions, and strictly before row $k$ in the construction of $\mathcal B$. This contradicts the choice of $k$. Hence $\mathcal B\cap\mathcal C=\mathcal A$.
\end{proof}

We recall two basic properties of information measures that will be used below.
\begin{property}[Submodularity and monotonicity]
\label{prop:entropy}
For all $\mathcal U,\mathcal V\subseteq\mathcal N$,
\begin{enumerate}
\item $H_{\mathcal U}+H_{\mathcal V}\ge H_{\mathcal U\cup\mathcal V}+H_{\mathcal U\cap\mathcal V}$;
\item $\mathcal U\subseteq\mathcal V$ implies $H_{\mathcal U}\le H_{\mathcal V}$.
\end{enumerate}
\end{property}
Consequently, $\mathcal U\subseteq\mathcal K\subseteq\mathcal V$ and $H_{\mathcal U}=H_{\mathcal V}$ imply $H_{\mathcal K}=H_{\mathcal U}$.

\Needspace{6\baselineskip}
\begin{lemma}\label{le:step1} Let $\mathcal A\subseteq\mathcal N$. If $\mathcal D_k\subseteq\mathcal A$ for some $k$, then $H_{\mathcal A}=H_{\mathcal A\cup\mathcal R_k}$.
	\end{lemma}
	\begin{proof}
		Since $\mathcal{D}_k \subseteq \mathcal{A}$, we have $\mathcal{A} \cup (\mathcal{D}_k \cup \mathcal{R}_k) = \mathcal{A} \cup \mathcal{R}_k$. 
		Furthermore, the intersection satisfies $\mathcal{D}_k \subseteq \mathcal{A} \cap (\mathcal{D}_k \cup \mathcal{R}_k) \subseteq \mathcal{D}_k \cup \mathcal{R}_k$.

		By Property~\ref{prop:entropy}(2) and the condition $H_{\mathcal{D}_k} = H_{\mathcal{D}_k \cup \mathcal{R}_k}$ (from \eqref{eqn:diagnew}), we have:
		\begin{equation}\label{eq:inter_entropy}
			H_{\mathcal{A} \cap (\mathcal{D}_k \cup \mathcal{R}_k)} = H_{\mathcal{D}_k \cup \mathcal{R}_k}.
		\end{equation}

		Applying the submodularity (Property~\ref{prop:entropy}(1)) to $\mathcal{A}$ and $\mathcal{D}_k \cup \mathcal{R}_k$, we obtain:
		\begin{equation}
			H_{\mathcal{A} \cup \mathcal{R}_k} + H_{\mathcal{A} \cap (\mathcal{D}_k \cup \mathcal{R}_k)} \le H_{\mathcal{A}} + H_{\mathcal{D}_k \cup \mathcal{R}_k}.
		\end{equation}

Substituting~\eqref{eq:inter_entropy} and cancelling the common term gives $H_{\mathcal A\cup\mathcal R_k}\le H_{\mathcal A}$. Monotonicity gives the reverse inequality, proving the claim.

		\end{proof}
\begin{lemma}
	\label{le:step2}
	Let $\mathcal A\subseteq\mathcal N$. If $\mathcal C_k\subseteq\mathcal A$ for some $k$, then $H_{\mathcal A}=H_{\mathcal A\cup\mathcal D_k}=H_{\mathcal A\cup\mathcal D_k\cup\mathcal R_k}$.
\end{lemma}
\begin{proof}
	By~\eqref{eqn:colnew}, we have $H_{\mathcal{C}_k} = H_{\mathcal{C}_k \cup \mathcal{D}_k}$. 
	Since $\mathcal{D}_k \subseteq \mathcal{C}_k \cup \mathcal{D}_k$, applying Lemma~\ref{le:step1} to the set $\mathcal{C}_k \cup \mathcal{D}_k$ yields $H_{\mathcal{C}_k \cup \mathcal{D}_k} = H_{\mathcal{C}_k \cup \mathcal{D}_k \cup \mathcal{R}_k}$. 
	Combining these two equalities gives:
	\begin{equation}\label{eq:col_full_ext}
		H_{\mathcal{C}_k} = H_{\mathcal{C}_k \cup \mathcal{D}_k \cup \mathcal{R}_k}.
	\end{equation}

	Next, since $\mathcal{C}_k \subseteq \mathcal{A}$, we apply the submodularity (Property~\ref{prop:entropy}(1)) to $\mathcal{A}$ and $\mathcal{C}_k \cup \mathcal{D}_k \cup \mathcal{R}_k$:
	\[
	H_{\mathcal{A} \cup \mathcal{D}_k \cup \mathcal{R}_k} + H_{\mathcal{A} \cap (\mathcal{C}_k \cup \mathcal{D}_k \cup \mathcal{R}_k)} \le H_{\mathcal{A}} + H_{\mathcal{C}_k \cup \mathcal{D}_k \cup \mathcal{R}_k}.
	\]
	Since $\mathcal{C}_k \subseteq \mathcal{A} \cap (\mathcal{C}_k \cup \mathcal{D}_k \cup \mathcal{R}_k) \subseteq \mathcal{C}_k \cup \mathcal{D}_k \cup \mathcal{R}_k$, monotonicity (Property~\ref{prop:entropy}(2)) and \eqref{eq:col_full_ext} imply:
	\[
	H_{\mathcal{A} \cap (\mathcal{C}_k \cup \mathcal{D}_k \cup \mathcal{R}_k)} = H_{\mathcal{C}_k \cup \mathcal{D}_k \cup \mathcal{R}_k}.
	\]
	Substituting this equality back into the submodularity inequality yields $H_{\mathcal{A} \cup \mathcal{D}_k \cup \mathcal{R}_k} \le H_{\mathcal{A}}$. 
	Combined with monotonicity $H_{\mathcal{A}} \le H_{\mathcal{A} \cup \mathcal{D}_k} \le H_{\mathcal{A} \cup \mathcal{D}_k \cup \mathcal{R}_k}$, we obtain:
	\[
	H_{\mathcal{A}} = H_{\mathcal{A}\cup\mathcal{D}_k} = H_{\mathcal{A}\cup\mathcal{D}_k \cup \mathcal{R}_k}.
	\]
\end{proof}

		Lemmas~\ref{le:step1} and~\ref{le:step2} directly yield Lemma~\ref{thm:closure}.

		\begin{lemma}\label{thm:closure}
			$H_{\mathcal A}=H_{\overline{\mathcal A}}$ for every $\mathcal A\subseteq\mathcal N$.
		\end{lemma}

		\Needspace{9\baselineskip}
\begin{lemma}[Implicit equalities]
			\label{thm:impliciteqs}
			The following equalities hold: 
			\begin{enumerate}
				\item $H_{\mathcal N}=H_{\mathcal N\setminus\{i\}}$ for every $i\in\mathcal N$.
				\item Let $\mathcal A\subseteq\mathcal N$ and $i,j\notin\mathcal A$, with $i\ne j$. If $i\in\overline{\mathcal A}$ or $j\in\overline{\mathcal A}$, then
				\[
				H_{\{i\} \cup \mathcal{A}} + H_{\{j\} \cup \mathcal{A}} - H_{\{i, j\} \cup \mathcal{A}} - H_{\mathcal{A}} = 0.
				\]
			\end{enumerate}
		\end{lemma}
		\begin{proof}
			For the first statement, note that $\overline{\mathcal{N}\setminus\{i\}} = \mathcal{N}$ for every $i \in \mathcal{N}$. The conclusion follows directly from Lemma~\ref{thm:closure}.

			For the second statement, assume without loss of generality that $i \in \overline{\mathcal{A}}$. 
			Since $\mathcal{A} \subseteq \{i\}\cup\mathcal{A} \subseteq \overline{\mathcal{A}}$, by monotonicity and $H_{\mathcal{A}}=H_{\overline{\mathcal{A}}}$, we have $H_{\{i\}\cup\mathcal{A}}=H_{\mathcal{A}}$. 
			Similarly, because $\{j\}\cup\mathcal{A} \subseteq \{i, j\} \cup\mathcal{A} \subseteq \{j\}\cup\overline{\mathcal{A}} \subseteq \overline{\{j\}\cup\mathcal{A}}$, and Lemma~\ref{thm:closure} implies $H_{\{j\}\cup\mathcal{A}}=H_{\overline{\{j\}\cup\mathcal{A}}}$, monotonicity forces $H_{\{j\}\cup\mathcal{A}}=H_{\{i, j\}\cup\mathcal{A}}$. 
			Substituting these two equalities into the expression yields:
			\[\begin{aligned}
				& H_{\{i\}\cup\mathcal{A}} + H_{\{j\}\cup\mathcal{A}} - H_{\{i, j\}\cup\mathcal{A}} - H_{\mathcal{A}} \\
				 & = H_{\mathcal{A}} + H_{\{j\}\cup\mathcal{A}} - H_{\{j\}\cup\mathcal{A}} - H_{\mathcal{A}} = 0,
			\end{aligned}
			\]
			which completes the proof.
		\end{proof}

		\Needspace{8\baselineskip}
		\begin{lemma}
			\label{thm:nonimpliciteqs}
			Let $\mathcal A\subseteq\mathcal N$ and let $i,j\notin\overline{\mathcal A}$ be distinct. Then the inequality 
			\[
			H_{\{i\}\cup\mathcal{A}} + H_{\{j\}\cup\mathcal{A}} - H_{\{i,j\}\cup\mathcal{A}} - H_{\mathcal{A}} \ge 0
			\]
			is not an implicit equality in system \eqref{sys:initial}.
		\end{lemma}
		\begin{proof}
			For a fixed index set $\mathcal{A}$ and $i, j \notin \overline{\mathcal{A}}$, we construct a specific evaluation function $H$ for entropy variables as follows:
			\begin{equation*}
				H_{\mathcal{K}}=\left\{
				\begin{aligned}
					0, &\quad \text{if } \mathcal{K}=\emptyset, \\
					1, &\quad \text{if } \emptyset \ne \mathcal{K}\subseteq\overline{\mathcal{A}}, \\
					2, &\quad \text{if } \mathcal{K}\not\subseteq\overline{\mathcal{A}}.
				\end{aligned}
				\right.
			\end{equation*}

			We now verify that this assignment satisfies all constraints in \eqref{sys:initial} while making the target inequality strictly positive:

			\begin{enumerate}[(1)]
				\item \textit{Closure and Boundary Constraints:} 
				Since $i, j \notin \overline{\mathcal{A}}$, we know $\overline{\mathcal{A}} \ne \mathcal{N}$ and $\mathcal{N} \setminus \{i'\} \not\subseteq \overline{\mathcal{A}}$ for any $i' \in \mathcal{N}$. Thus, $H_{\mathcal{N}} = H_{\mathcal{N} \setminus \{i'\}} = 2$, satisfying $H_{\mathcal{N}} = H_{\mathcal{N}\setminus\{i'\}}$ \eqref{ineq:twonew}.

				\item \textit{Row and Column Criterion Constraints:} 
				Since $\overline{\mathcal{A}}$ is a closed set, if $\mathcal{D}_k \subseteq \overline{\mathcal{A}}$, then by definition $\mathcal{D}_k \cup \mathcal{R}_k \subseteq \overline{\mathcal{A}}$. The common value of $H_{\mathcal D_k}$ and $H_{\mathcal D_k\cup\mathcal R_k}$ is $1$ when $\mathcal D_k\subseteq\overline{\mathcal A}$, and $2$ otherwise. In either case, $H_{\mathcal{D}_k} = H_{\mathcal{D}_k \cup \mathcal{R}_k}$, satisfying \eqref{eqn:diagnew}. The same argument applies to \eqref{eqn:colnew}.

				\item \textit{Submodularity Inequalities:} 
				For any index set $\mathcal{K}$ and $i', j' \notin \mathcal{K}$, consider $\Delta = H_{\{i'\}\cup\mathcal{K}} + H_{\{j'\}\cup\mathcal{K}} - H_{\{i',j'\}\cup\mathcal{K}} - H_{\mathcal{K}}$.
				Evaluating $\Delta$ in all cases gives
				\begin{equation*}
					\Delta = \left\{
					\begin{aligned}
						1, &\quad \text{if } i', j' \notin \overline{\mathcal{A}} \text{ and } \emptyset \ne \mathcal{K} \subseteq \overline{\mathcal{A}}, \\
						1, &\quad \text{if } |\{i', j'\} \cap \overline{\mathcal{A}}| = 1 \text{ and } \mathcal{K} = \emptyset, \\
						1, &\quad \text{if } \{i', j'\} \subseteq \overline{\mathcal{A}} \text{ and } \mathcal{K} = \emptyset, \\
						2, &\quad \text{if } i', j' \notin \overline{\mathcal{A}} \text{ and } \mathcal{K} = \emptyset, \\
						0, &\quad \text{otherwise.}
					\end{aligned}
					\right.
				\end{equation*}
				In all cases, $\Delta \ge 0$, so all inequalities in \eqref{ineq:fournew} are satisfied.
			\end{enumerate}

			Finally, for the specific choice of $\mathcal{A}$ and $i, j \notin \overline{\mathcal{A}}$:
			\begin{itemize}
				\item If $\mathcal{A} \neq \emptyset$, then $\emptyset \neq \mathcal{A} \subseteq \overline{\mathcal{A}}$, which falls into the first case ($\Delta = 1 > 0$).
				\item If $\mathcal{A} = \emptyset$, it falls into the fourth case ($\Delta = 2 > 0$).
			\end{itemize}
			In both cases, $H_{\{i\}\cup\mathcal{A}} + H_{\{j\}\cup\mathcal{A}} - H_{\{i,j\}\cup\mathcal{A}} - H_{\mathcal{A}} > 0$. Thus the target inequality is strict at a feasible point and is not an implicit equality.
		\end{proof}

For the equivalence proof, Lemma~\ref{thm:closure} identifies $H_{\mathcal B}$ with $H_{\overline{\mathcal B}}$ for every $\mathcal B\subseteq\mathcal N$. Under this substitution, the repair equalities and the two-term elemental inequalities become identities. Lemmas~\ref{thm:impliciteqs} and~\ref{thm:nonimpliciteqs} identify exactly which remaining elemental inequalities are implicit equalities. The nontrivial constraints can be written as
\[
H_{\overline{\mathcal A\cup\{i\}}}+H_{\overline{\mathcal A\cup\{j\}}}
-H_{\overline{\mathcal A\cup\{i,j\}}}-H_{\mathcal A}\ge0,
\quad \mathcal A\in\mathcal{CS},\quad i,j\notin\mathcal A,\ i\ne j.
\]
Write $\mathcal B=\overline{\mathcal A\cup\{i\}}$ and $\mathcal C=\overline{\mathcal A\cup\{j\}}$. These two closed sets are \emph{comparable} if one contains the other. If $\mathcal B\subseteq\mathcal C$, then $\overline{\mathcal B\cup\mathcal C}=\mathcal C$, and two terms cancel. The reverse inclusion is covered by exchanging $i$ and $j$. If the sets are incomparable, all four coordinates remain. We therefore state the equivalent intermediate system in these two forms.

\begin{construction}\label{thm:eliminatedsys}
For each closed $\mathcal A$ and distinct $i,j\in\mathcal N\setminus\mathcal A$, put $\mathcal B=\overline{\mathcal A\cup\{i\}}$ and $\mathcal C=\overline{\mathcal A\cup\{j\}}$. The distinct constraints are
\begin{subequations}\label{sys:elimedsystem}\label{sys:elimedsystemseperate}
\begin{empheq}[left=\empheqlbrace]{align}
&H_{\mathcal B}+H_{\mathcal C}-H_{\overline{\mathcal B\cup\mathcal C}}-H_{\mathcal A}\ge0,
\quad \mathcal B\not\subseteq\mathcal C,\ \mathcal C\not\subseteq\mathcal B,\label{ineq:eliminatedfour}\\
&H_{\mathcal B}-H_{\mathcal A}\ge0,\quad\mathcal B\subseteq\mathcal C.\label{ineq:eliminatedtwo}
\end{empheq}
\end{subequations}
\end{construction}

\begin{figure}[ht]

\centering
\includegraphics[width=0.4\linewidth]{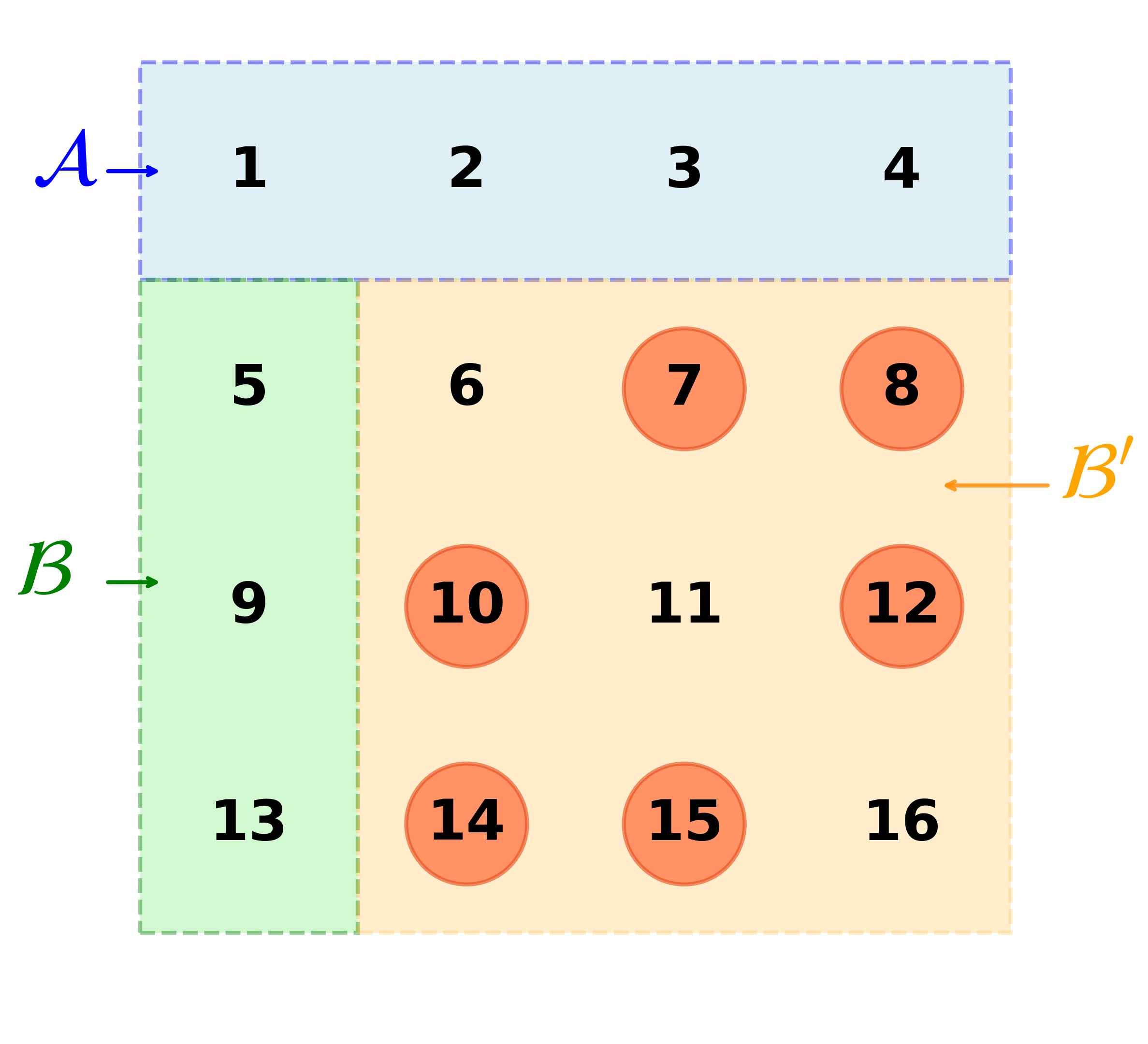}
\caption{Partition outside the selected full rows: unrestricted entries in $\mathcal B$ and column-constrained entries in $\mathcal B'$.}
\label{fig:closedset}
\end{figure}

\begin{theorem}[Number of closed sets]
\label{thm:numofclosures}
For $n\ge2$, the total number of closed sets is
\begin{equation}\label{eq:closed-count}
\textsf{num\_CS}=1+\sum_{k=0}^{n-2}\binom nk 2^{k(n-k)}(2^{n-k-1}-1)^{n-k}.
\end{equation}
Moreover, $\textsf{num\_CS}=O(2^{n^2-n})$ as $n\to\infty$.
\end{theorem}
\begin{proof}
A closed set consists of some full rows and additional non-diagonal entries. It cannot contain exactly $n-1$ full rows: the remaining non-diagonal column would be complete, and the column criterion would fill the last row. The unique set with $n$ full rows is $\mathcal N$, accounting for the leading $1$.

Fix $k\le n-2$ full rows; there are $\binom nk$ choices. Let $\mathcal I$ be their row indices and let $\mathcal I'$ be the complement. Outside the full rows, partition the entries into $\mathcal B$ (columns in $\mathcal I$) and $\mathcal B'$ (columns in $\mathcal I'$), as in Fig.~\ref{fig:closedset}. The $k(n-k)$ entries of $\mathcal B$ are unrestricted: completing a column there can only add a row that is already full. In each of the $n-k$ columns of $\mathcal B'$, the diagonal entry must be absent, and the $n-k-1$ remaining off-diagonal entries cannot all be chosen. Thus each such column has $2^{n-k-1}-1$ choices. These column choices and the choices in $\mathcal B$ are independent. This proves \eqref{eq:closed-count}.

For the asymptotic bound, replace $2^{n-k-1}-1$ by $2^{n-k-1}$. The $k$th summand is then bounded by $\binom nk2^{(n-k)(n-1)}$. Consequently,
\[
\textsf{num\_CS}\le1+2^{n^2-n}(1+2^{-(n-1)})^n,
\]
and the final factor is bounded for $n\ge2$.
\end{proof}

\begin{corollary}[Size of System~\ref{thm:eliminatedsys}]\label{cor:system2-size}
System~\ref{thm:eliminatedsys} has $\textsf{num\_CS}-1$ entropy coordinates, one for each nonempty closed set. Its number $\textsf{num\_ES}$ of distinct inequalities satisfies
\begin{equation}\label{eq:coarse-bound}
\textsf{num\_ES}\le\sum_{\mathcal A\in\mathcal{CS}}\binom{n^2-|\mathcal A|}{2}
\le\textsf{num\_CS}\binom{n^2}{2}=O(2^{n^2-n}n^4).
\end{equation}
\end{corollary}
\begin{proof}
The closure substitution leaves one coordinate for each closed set, except the fixed coordinate $H_\emptyset=0$. For a fixed closed $\mathcal A$, each unordered pair of indices outside $\mathcal A$ gives at most one constraint before repeated expressions are identified. Summing these candidate counts and applying Theorem~\ref{thm:numofclosures} proves the bound.
\end{proof}
The next subsection defines System~\ref{const:final} from the closed-set structure and proves its equivalence with System~\ref{const:initial}.

\subsection{Direct definition and equivalence of System~\ref{const:final}}
	\label{sec:redundantineqs}

\begin{define}[Closed-set graph and inseparable edges]\label{def:closed-graph}
The closed-set graph has a node for every $\mathcal A\in\mathcal{CS}$ and a directed edge $\mathcal A\to\mathcal B$ whenever $\mathcal B=\overline{\mathcal A\cup\{i\}}$ for some $i\notin\mathcal A$. An edge is \emph{inseparable} if no other directed path joins its endpoints, and \emph{separable} otherwise.
\end{define}
Equivalently, $\mathcal A\to\mathcal B$ is inseparable precisely when no closed set $\mathcal K$ satisfies $\mathcal A\subsetneq\mathcal K\subsetneq\mathcal B$. To see this, an alternative path supplies an intermediate closed node. Conversely, given such a $\mathcal K$, adjoining its elements one at a time gives a path from $\mathcal A$ to $\mathcal K$. Adjoining an index that generates $\mathcal B$ then completes an alternative path.

Let $\mathcal E(\mathcal A)$ be the set of distinct targets of outgoing inseparable edges from $\mathcal A$. Thus $\mathcal E(\mathcal A)$ consists of the minimal proper closed supersets of $\mathcal A$.

	\begin{figure}[ht]
		\centering
		\includegraphics[width=.7\linewidth]{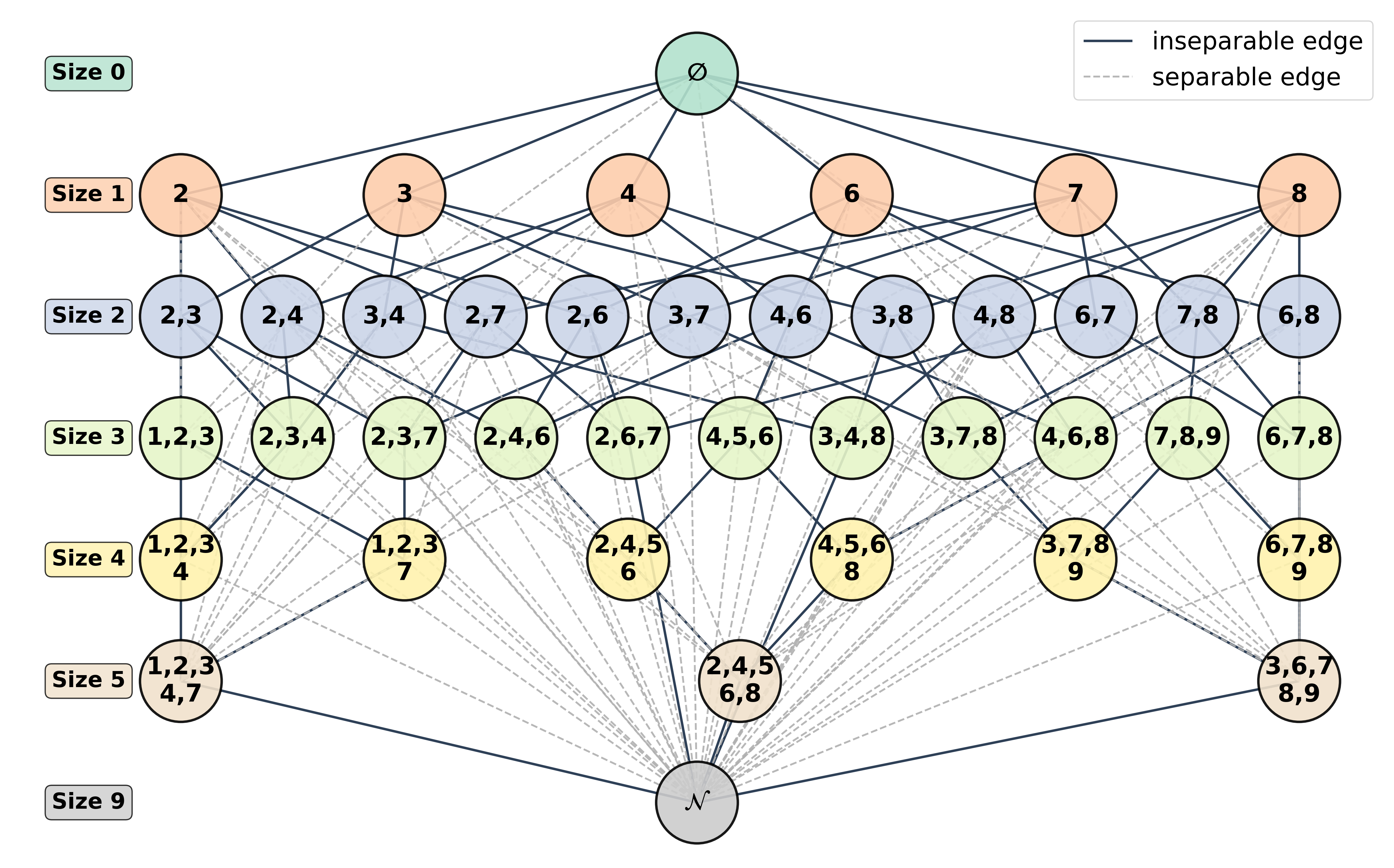}
		\caption{The closed-set graph for $n=3$. Each edge represents a single-index extension followed by closure. Solid and dashed edges are inseparable and separable, respectively.}
		\label{fig:graphfull}
	\end{figure}

Fig.~\ref{fig:graphfull} shows how the graph represents the constraints in System~\ref{thm:eliminatedsys}. In a four-term inequality, the two edges from $\mathcal A$ end at $\mathcal B$ and $\mathcal C$, and the fourth node is their join $\overline{\mathcal B\cup\mathcal C}$. For example, the edges $\emptyset\to\{2\}$ and $\emptyset\to\{4\}$ give $H_{\{2\}}+H_{\{4\}}-H_{\{2,4\}}\ge0$. A two-term inequality uses a single edge from $\mathcal A$ to $\mathcal B$. We call $\mathcal A$ the \emph{initial node} of the inequality. The distinction between solid and dashed edges will specify which constraints occur in the directly defined final system.

The repair dependencies in System~\ref{const:initial} determine the closure operation, the closed sets, and their inseparable edges. These objects specify the following system directly.

\Needspace{8\baselineskip}
\begin{construction}\label{const:final}
For each closed $\mathcal A\ne\mathcal N$, the constraints are
\begin{subequations}\label{sys:final}
\begin{empheq}[left=\empheqlbrace]{align}
&H_{\mathcal P}+H_{\mathcal Q}-H_{\overline{\mathcal P\cup\mathcal Q}}-H_{\mathcal A}\ge0,
\quad \mathcal P,\mathcal Q\in\mathcal E(\mathcal A),\ \mathcal P\ne\mathcal Q,\label{sys:final-four}\\
&H_{\mathcal P}-H_{\mathcal A}\ge0,\quad\mathcal E(\mathcal A)=\{\mathcal P\}.\label{sys:final-two}
\end{empheq}
\end{subequations}
Each unordered pair in~\eqref{sys:final-four} contributes once, and each distinct inequality is listed once.
\end{construction}

The top set $\mathcal N$ has no outgoing edge and contributes no inequality. System~\ref{const:final} is specified entirely by the closure structure induced by the repair dependencies of System~\ref{const:initial}. The implicit equalities established above and the redundancy criteria below prove equivalence of the two systems. The definition itself does not depend on an iterative search for redundant constraints. These sufficient criteria concern the excluded constraints; Theorem~\ref{thm:general-irredundancy} below separately proves that none of the retained inequalities is redundant.

\Needspace{13\baselineskip}
\begin{theorem}[Redundancy Identification]
	\label{thm:redundantineqs}
	Let $\mathcal A$ be closed. For constraints belonging to system~\eqref{sys:elimedsystemseperate}, the following sufficient criteria hold:
	\begin{enumerate}[(1)]
		\item The four-term inequality $H_{\overline{\{i\}\cup\Acal}} + H_{\overline{\{j\}\cup\Acal}} - H_{\overline{\{i,j\}\cup\Acal}} - H_{\Acal} \ge 0$ is redundant if at least one of the edges $\Acal \rightarrow \overline{\{i\}\cup\Acal}$ or $\Acal \rightarrow \overline{\{j\}\cup\Acal}$ is separable. \label{item:type1}
		\item The two-term inequality $H_{\overline{\{i\}\cup\Acal}} - H_{\Acal} \ge 0$ is redundant if the edge $\Acal \rightarrow \overline{\{i\}\cup\Acal}$ is separable. \label{item:type2}
		\item When the edge $\Acal \rightarrow \overline{\{i\}\cup\Acal}$ is inseparable, the two-term inequality $H_{\overline{\{i\}\cup\Acal}} - H_{\Acal} \ge 0$ is redundant if there exists another distinct inseparable edge originating from $\Acal$. \label{item:type3}
	\end{enumerate}
\end{theorem}

\begin{proof}
	We first prove~(\ref{item:type1}). 
	Without loss of generality, assume that the edge $\Acal \rightarrow \overline{\{i\}\cup\Acal}$ is separable. 
	Then, there exists an inseparable edge $\Acal \rightarrow \overline{\{k\}\cup\Acal}$ for some $k \in \Ncal \setminus \Acal$ such that $\overline{\{k\}\cup\Acal} \subsetneqq \overline{\{i\}\cup\Acal}$.
	Note that $\overline{\{i,k\}\cup\Acal} = \overline{\{i\}\cup\Acal}$ and $\overline{\{i,j,k\}\cup\Acal} = \overline{\{i,j\}\cup\Acal}$.

	If $\overline{\{k\}\cup\Acal} \not\subseteq \overline{\{j\}\cup\Acal}$ and $\overline{\{j\}\cup\Acal} \not\subseteq \overline{\{k\}\cup\Acal}$, summing the two four-term inequalities:
	\[
	H_{\overline{\{k\}\cup\Acal}} + H_{\overline{\{j\}\cup\Acal}} - H_{\overline{\{k,j\}\cup\Acal}} - H_{\Acal} \ge 0,
	\]
	and
	\[
	H_{\overline{\{j,k\}\cup\Acal}} + H_{\overline{\{i,k\}\cup\Acal}} - H_{\overline{\{i,j,k\}\cup\Acal}} - H_{\overline{\{k\}\cup\Acal}} \ge 0,
	\]
	directly yields $H_{\overline{\{i\}\cup\Acal}} + H_{\overline{\{j\}\cup\Acal}} - H_{\overline{\{i,j\}\cup\Acal}} - H_{\Acal} \ge 0$.
	If $\overline{\{k\}\cup\Acal} \subseteq \overline{\{j\}\cup\Acal}$, the first inequality simplifies into a two-term inequality $H_{\overline{\{k\}\cup\Acal}} - H_{\Acal} \ge 0$, and summing it with the second four-term inequality produces the identical result. 
	Hence, Statement~(\ref{item:type1}) holds.

	\medskip
	Next, we prove~(\ref{item:type2}). 
	When $\Acal \rightarrow \overline{\{i\}\cup\Acal}$ is separable, there exists an inseparable edge $\Acal \rightarrow \overline{\{k\}\cup\Acal}$ with $k \in \Ncal \setminus \Acal$ such that $\overline{\{k\}\cup\Acal} \subsetneqq \overline{\{i\}\cup\Acal}$.
	Recall that the two-term inequality $H_{\overline{\{i\}\cup\Acal}} - H_{\Acal} \ge 0$ degenerates from a four-term inequality for some $j \in \Ncal \setminus \Acal$ satisfying $\overline{\{i\}\cup\Acal} \subseteq \overline{\{j\}\cup\Acal}$.
	Since $\overline{\{i\}\cup\Acal} = \overline{\{i,k\}\cup\Acal} \subseteq \overline{\{j,k\}\cup\Acal}$, both $H_{\overline{\{k\}\cup\Acal}} - H_{\Acal} \ge 0$ and $H_{\overline{\{i\}\cup\Acal}} - H_{\overline{\{k\}\cup\Acal}} \ge 0$ belong to the set of valid two-term inequalities in \eqref{sys:elimedsystemseperate}.
	Summing these two inequalities gives $H_{\overline{\{i\}\cup\Acal}} - H_{\Acal} \ge 0$, establishing its redundancy.

	\medskip
	Lastly, we prove~(\ref{item:type3}). 
	Suppose there exists another distinct inseparable edge $\Acal \rightarrow \overline{\{k\}\cup\Acal}$ originating from $\Acal$ for some $k \neq i$.
	Similarly, $H_{\overline{\{i\}\cup\Acal}} - H_{\Acal} \ge 0$ corresponds to the condition $\overline{\{i\}\cup\Acal} \subseteq \overline{\{j\}\cup\Acal}$ for some $j$.
	Because $\overline{\{i,k\}\cup\Acal} \subseteq \overline{\{j,k\}\cup\Acal}$, the inequality $H_{\overline{\{i,k\}\cup\Acal}} - H_{\overline{\{k\}\cup\Acal}} \ge 0$ is a valid two-term inequality in system \eqref{sys:elimedsystemseperate}.
	Adding it to the four-term inequality $H_{\overline{\{i\}\cup\Acal}} + H_{\overline{\{k\}\cup\Acal}} - H_{\overline{\{i,k\}\cup\Acal}} - H_{\Acal} \ge 0$ eliminates the intermediate terms $H_{\overline{\{k\}\cup\Acal}}$ and $H_{\overline{\{i,k\}\cup\Acal}}$, leaving $H_{\overline{\{i\}\cup\Acal}} - H_{\Acal} \ge 0$.
	Thus, Statement~(\ref{item:type3}) is proven.
\end{proof}

Example~\ref{ex:redineqs} illustrates the three redundancy criteria in system \eqref{sys:elimedsystemseperate}.

	\begin{example}[Redundant Inequalities]
		\label{ex:redineqs}
	 The following redundancies are observed in \eqref{sys:elimedsystemseperate}:
		\begin{itemize}
			\item[\textbf{1)}] The inequality $H_{\{7,8,9\}} + H_{\{2\}} - H_{\Ncal} - H_{\emptyset} \ge 0$ is redundant, being derivable from the sum of $H_{\{2\}} + H_{\{8\}} - H_{\{2,4,5,6,8\}} - H_{\emptyset} \ge 0$ and $H_{\{2,4,5,6,8\}} + H_{\{7,8,9\}} - H_{\Ncal} - H_{\{8\}} \ge 0$.
			The edge $\emptyset \rightarrow \{7,8,9\} = \overline{\{9\}}$  is separable, as witnessed by the alternative path $\emptyset \rightarrow \{7\} \rightarrow \{7, 8\} \rightarrow \{7, 8, 9\}$.
			\item[\textbf{2)}] The inequality $H_{\{1,2,3,4\}} - H_{\{2,4\}} \ge 0$ is redundant, since it is obtained by adding $H_{\{1,2,3,4\}} - H_{\{2,3,4\}} \ge 0$ and $H_{\{2,3,4\}} - H_{\{2,4\}} \ge 0$.
			The edge $\{2,4\} \rightarrow \{1,2,3,4\} = \overline{\{1,2,4\}}$  is separable, as the alternative path  $\{2,4\} \rightarrow \{2,3,4\} \rightarrow \{1,2,3,4\}$ exists.
			\item[\textbf{3)}] The inequality $H_{\{2\}} - H_{\emptyset} \ge 0$ is redundant, as it follows from $H_{\{2\}} + H_{\{4\}} - H_{\{2,4\}} - H_{\emptyset} \ge 0$ and $H_{\{2,4\}} - H_{\{4\}} \ge 0$.
			 In this case, the edge $\emptyset \rightarrow \{2\}$ is  inseparable; however, its initial node $\emptyset$ also admits another inseparable edge $\emptyset \rightarrow \{4\}$.
		\end{itemize}
		The graph structures corresponding to these three cases are shown in Fig.~\ref{fig:redundantineqs}.

		\begin{figure}[H]
			\centering
			\subfloat[\label{fig:redundant1}]{\includegraphics[width=0.33\linewidth]{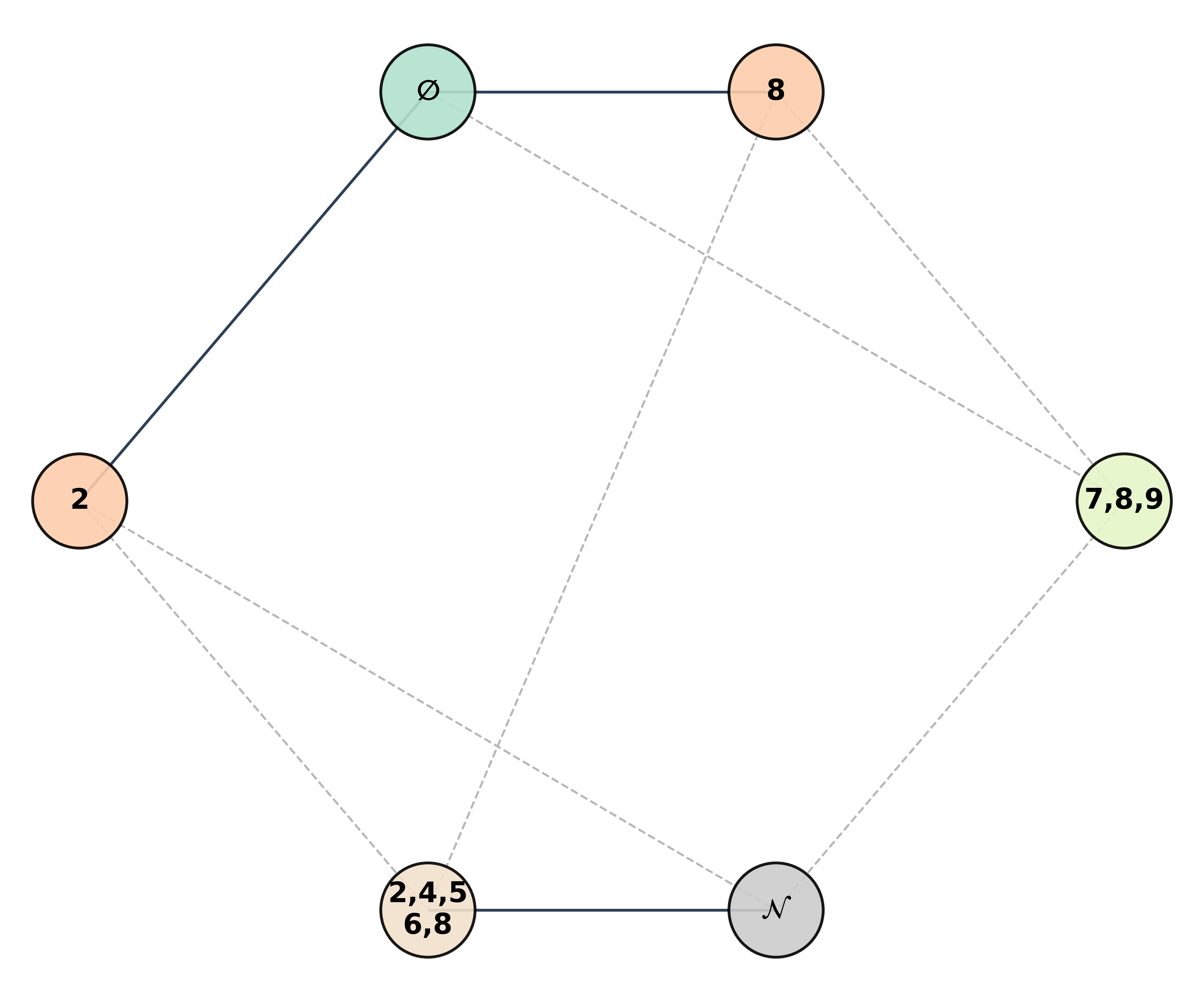}}
			\hfill
			\subfloat[\label{fig:redundant2}]{\includegraphics[width=0.33\linewidth]{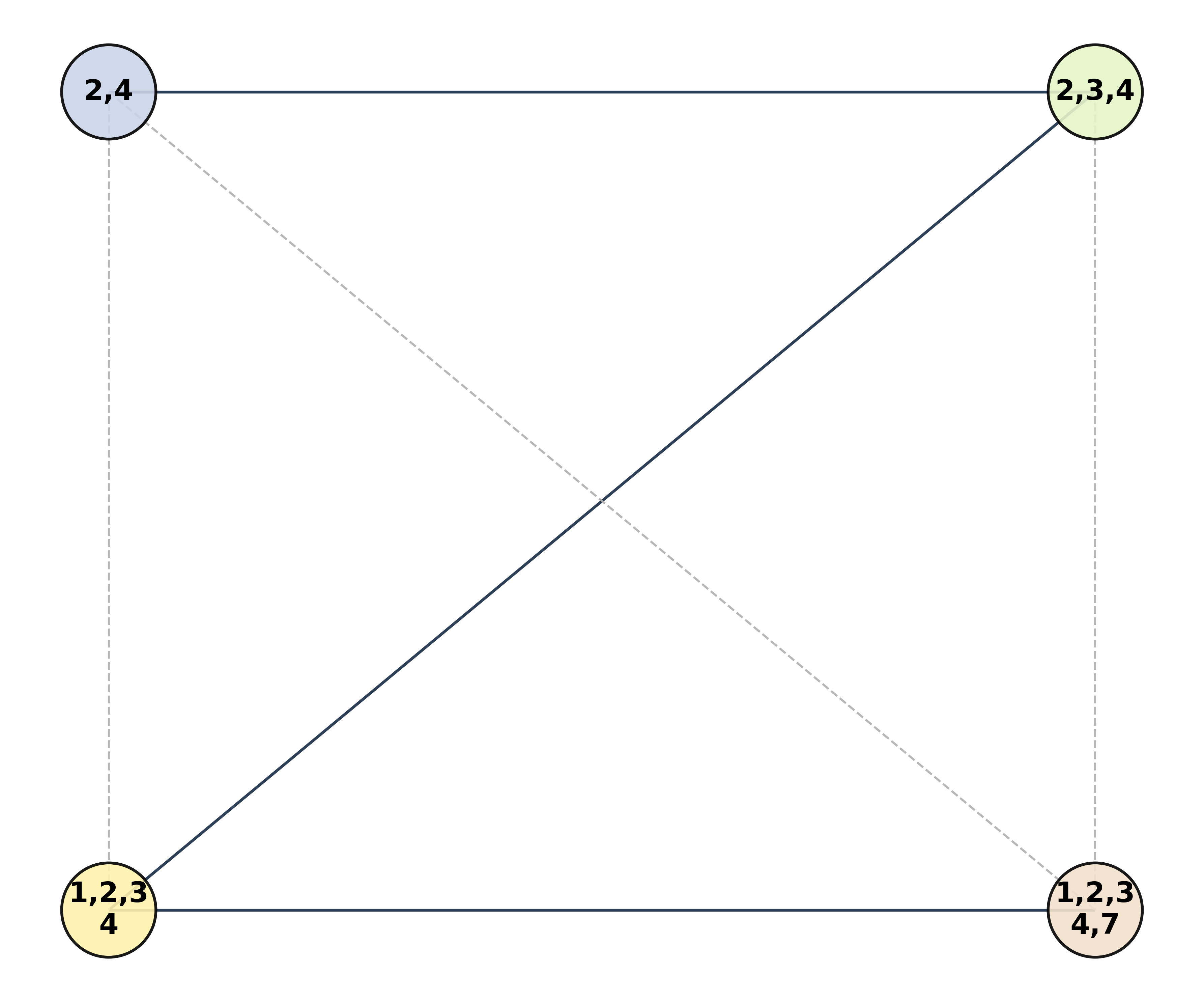}}
			\hfill
			\subfloat[\label{fig:redundant3}]{\includegraphics[width=0.33\linewidth]{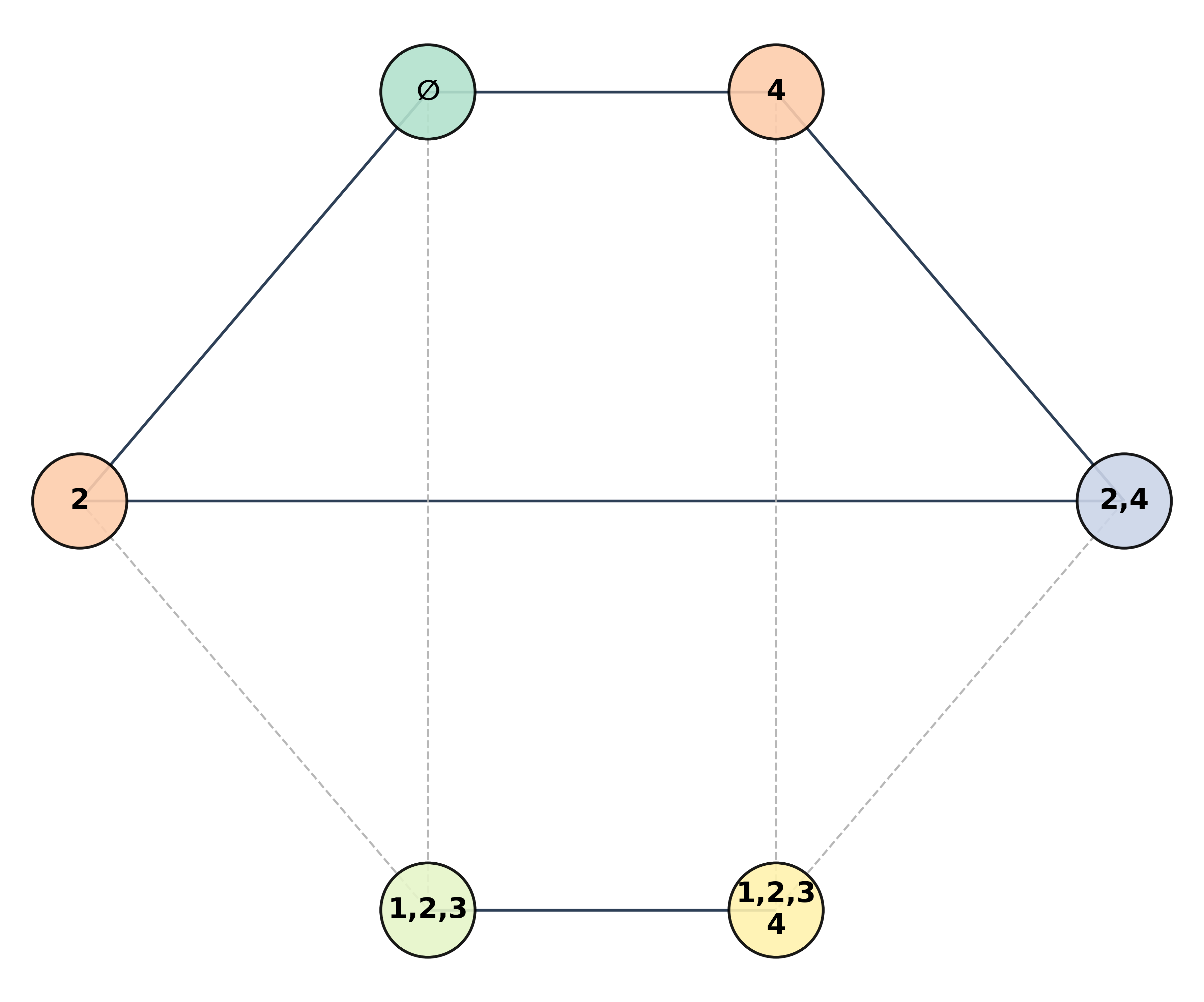}}
			\caption{Three representative classes of redundant inequalities.}
			\label{fig:redundantineqs}
		\end{figure}
	\end{example}

\Needspace{12\baselineskip}
\begin{corollary}[Equivalence of Systems~\ref{const:initial} and~\ref{const:final}]\label{cor:equiv-final}
Let $h=(H_{\mathcal B})_{\emptyset\ne\mathcal B\subseteq\mathcal N}$ be an entropy-coordinate vector and let $g=(g_{\mathcal C})_{\emptyset\ne\mathcal C\in\mathcal{CS}}$ be a vector indexed by nonempty closed sets. In System~\ref{const:final}, interpret $H_{\mathcal C}$ as $g_{\mathcal C}$. Set $H_\emptyset=g_\emptyset=0$. Then $h$ satisfies System~\ref{const:initial} if and only if there exists a vector $g$ satisfying System~\ref{const:final} such that
\begin{equation}\label{eq:closure-lift}
H_{\mathcal B}=g_{\overline{\mathcal B}}\qquad\text{for every }\mathcal B\subseteq\mathcal N.
\end{equation}
The vector $g$ is unique and is obtained by restricting $h$ to the nonempty closed sets.
\end{corollary}
\begin{proof}
For a vector satisfying System~\ref{const:initial}, Lemma~\ref{thm:closure} gives $H_{\mathcal B}=H_{\overline{\mathcal B}}$ for every $\mathcal B\subseteq\mathcal N$. Restriction to closed coordinates therefore gives~\eqref{eq:closure-lift}, and the preceding reduction yields System~\ref{thm:eliminatedsys}. Conversely, a lift defined by~\eqref{eq:closure-lift} satisfies the repair equalities automatically. Each elemental inequality becomes either an identity or a constraint in System~\ref{thm:eliminatedsys}. It remains to prove equivalence of Systems~\ref{thm:eliminatedsys} and~\ref{const:final}.

System~\ref{thm:eliminatedsys} implies the four-term constraints of System~\ref{const:final}, since each pair of distinct cover targets is generated by a pair of indices. Its equivalence with System~\ref{const:initial} also gives monotonicity, and hence the two-term constraints of System~\ref{const:final}.

Conversely, assume System~\ref{const:final}. We first prove cover monotonicity by downward induction on the initial closed set $\mathcal A$. If $\mathcal E(\mathcal A)=\{\mathcal P\}$, then $H_{\mathcal P}\ge H_{\mathcal A}$ is imposed explicitly. Otherwise, for a given $\mathcal P\in\mathcal E(\mathcal A)$, choose a distinct $\mathcal Q\in\mathcal E(\mathcal A)$. The induction hypothesis gives monotonicity along a chain of covers from $\mathcal Q$ to $\overline{\mathcal P\cup\mathcal Q}$, all of whose initial nodes strictly contain $\mathcal A$. The corresponding four-term constraint then gives
\[
H_{\mathcal P}-H_{\mathcal A}\ge H_{\overline{\mathcal P\cup\mathcal Q}}-H_{\mathcal Q}\ge0.
\]
Summing cover inequalities gives monotonicity for every pair of comparable closed sets, so all two-term constraints of System~\ref{thm:eliminatedsys} hold.

For a four-term constraint with initial node $\mathcal A$, targets $\mathcal B,\mathcal C$, and join $\mathcal J=\overline{\mathcal B\cup\mathcal C}$, use induction on the lexicographically ordered pair
\[
\bigl(|\mathcal J\setminus\mathcal A|,\ |\mathcal B\setminus\mathcal A|+|\mathcal C\setminus\mathcal A|\bigr).
\]
If both outgoing edges are inseparable, the constraint belongs to System~\ref{const:final}. Otherwise, apply the decomposition in Theorem~\ref{thm:redundantineqs}(\ref{item:type1}), using a cover target $\mathcal D\subsetneq\mathcal B$. The first summand has targets $\mathcal D,\mathcal C$ and join contained in $\mathcal J$, so its induction pair decreases. The second has initial node $\mathcal D$ and join $\mathcal J$, so its first induction component decreases. Any two-term summand is already covered by monotonicity. Thus the decomposition proves the target inequality from constraints with smaller induction pairs. All constraints of System~\ref{thm:eliminatedsys} follow.

Finally, $\overline{\mathcal C}=\mathcal C$ for a closed set $\mathcal C$, so~\eqref{eq:closure-lift} forces $g_{\mathcal C}=H_{\mathcal C}$ and proves uniqueness.
\end{proof}

\subsection{Irredundancy and full dimensionality}
\label{sec:general-irredundancy}
We now prove that every distinct inequality retained in System~\ref{const:final} is necessary for every $n\ge2$. This is a separate question from the equivalence established in Corollary~\ref{cor:equiv-final}. The proof uses only the closed sets of Definition~\ref{def:closure}, their cover targets in Definition~\ref{def:closed-graph}, and the two forms in~\eqref{sys:final}; it does not depend on the construction in Section~\ref{sec:edge} or on computations for particular values of $n$.

Let $D=|\mathcal{CS}|-1$ and work in the real coordinate space
\[
h=(h_{\mathcal C})_{\emptyset\ne\mathcal C\in\mathcal{CS}}\in\mathbb R^D,
\qquad h_\emptyset=0.
\]
Here $h_{\mathcal C}$ is the coordinate denoted by $H_{\mathcal C}$ in System~\ref{const:final}; a vector in this space need not be an entropy vector of actual random variables. No further equalities or coordinate lower bounds are imposed when testing whether one defining inequality follows from the others. For closed sets, write $\mathcal P\vee\mathcal Q=\overline{\mathcal P\cup\mathcal Q}$. Denote the distinct defining linear forms by $\mathscr R$ and their feasible cone by
\begin{equation}\label{eq:final-cone}
\mathscr C_3=\{h\in\mathbb R^D:r(h)\ge0\text{ for all }r\in\mathscr R\}.
\end{equation}
For a row whose initial node is $\mathcal A$, set $\beta(r)=\mathcal A$. Thus its form is either
\[
r(h)=h_{\mathcal P}+h_{\mathcal Q}-h_{\mathcal P\vee\mathcal Q}-h_{\mathcal A}
\quad\text{or}\quad
r(h)=h_{\mathcal P}-h_{\mathcal A},
\]
with the cover conditions specified in System~\ref{const:final}. Distinct covers $\mathcal P,\mathcal Q$ of $\mathcal A$ are incomparable: a strict inclusion between them would give a closed set between $\mathcal A$ and one of its covers. Consequently,
\[
\mathcal A\subsetneq\mathcal P,\mathcal Q\subsetneq\mathcal P\vee\mathcal Q,
\]
so the four coordinates in a four-term row are distinct.

\begin{lemma}[Coefficient rows and minimal initial nodes]\label{lem:row-support}
For $r\in\mathscr R$, let $r^\circ\in\mathbb R^D$ be its coefficient row, and let $\widetilde r\in\mathbb R^{\mathcal{CS}}$ be the row obtained by restoring the coefficient of $h_\emptyset$ from its displayed form. Then:
\begin{enumerate}[(1)]
\item Any relation $r^\circ=\sum_s\lambda_s s^\circ$ also satisfies $\widetilde r=\sum_s\lambda_s\widetilde s$.
\item If $w=\sum_{s\in\mathscr S}\lambda_s\widetilde s$ with all $\lambda_s>0$, and $\mathcal D$ is an inclusion-minimal member of $\{\beta(s):s\in\mathscr S\}$, then
\begin{equation}\label{eq:minimal-bottom}
w_{\mathcal D}=-\sum_{\substack{s\in\mathscr S\\\beta(s)=\mathcal D}}\lambda_s<0.
\end{equation}
\end{enumerate}
\end{lemma}
\begin{proof}
Every complete coefficient row has coefficient sum zero. Hence its empty-set coefficient is the negative sum of its nonempty coefficients, which proves (1). Restoring this column is only an algebraic device; $h_\emptyset$ remains fixed at zero.

Each complete row has coefficient $-1$ at its initial node, and every other nonzero coordinate strictly contains that node. If a row with initial node different from $\mathcal D$ contributed to the $\mathcal D$ column, its initial node would be a strict subset of $\mathcal D$, contradicting minimality. Only rows with initial node $\mathcal D$ contribute, proving (2).
\end{proof}

\begin{lemma}[Feasible test vectors]\label{lem:principal-tests}
For a closed set $\mathcal K$, define
\begin{equation}\label{eq:principal-test}
v^{\mathcal K}_{\mathcal C}=\begin{cases}
0,&\mathcal C\subseteq\mathcal K,\\
1,&\mathcal C\not\subseteq\mathcal K.
\end{cases}
\end{equation}
Then $v^{\mathcal K}\in\mathscr C_3$. For a four-term row with initial node $\mathcal A$ and cover targets $\mathcal P,\mathcal Q$,
\begin{equation}\label{eq:principal-slack}
r(v^{\mathcal K})=\begin{cases}
1,&\mathcal A\subseteq\mathcal K,\quad
\mathcal P\not\subseteq\mathcal K,\quad\mathcal Q\not\subseteq\mathcal K,\\
0,&\text{otherwise}.
\end{cases}
\end{equation}
\end{lemma}
\begin{proof}
If $\mathcal A\not\subseteq\mathcal K$, all four coordinates equal one, giving zero slack. If $\mathcal A\subseteq\mathcal K$, its coordinate is zero. Since $\mathcal K$ is closed, $\mathcal P\vee\mathcal Q\subseteq\mathcal K$ holds exactly when both $\mathcal P$ and $\mathcal Q$ are subsets of $\mathcal K$. Evaluating the remaining three coordinates gives~\eqref{eq:principal-slack}. For a two-term row, $\mathcal A\subsetneq\mathcal P$ implies $v^{\mathcal K}_{\mathcal P}\ge v^{\mathcal K}_{\mathcal A}$. Finally, $v^{\mathcal K}_\emptyset=0$, as required.
\end{proof}

\Needspace{8\baselineskip}
\begin{theorem}[Irredundancy of System~\ref{const:final}]\label{thm:general-irredundancy}
For every $n\ge2$, each distinct inequality in System~\ref{const:final} is not implied by the remaining inequalities in the nonempty closed-set coordinate space $\mathbb R^D$.
\end{theorem}
\begin{proof}
Fix a target row $r$. If it were redundant, the homogeneous Farkas lemma~\cite[Theorem~1.11]{YangFarkas} would express $r^\circ$ as a nonnegative combination of the other coefficient rows. By Lemma~\ref{lem:row-support}(1), this relation lifts to
\begin{equation}\label{eq:irredundancy-combination}
\widetilde r=\sum_{s\in\mathscr S}\lambda_s\widetilde s,
\qquad \mathscr S\subseteq\mathscr R\setminus\{r\},\quad\lambda_s>0,
\end{equation}
after zero weights are discarded. The support $\mathscr S$ is nonempty because $r$ is nonzero.

First suppose that $r(h)=h_{\mathcal P}+h_{\mathcal Q}-h_{\mathcal J}-h_{\mathcal A}$, where $\mathcal J=\mathcal P\vee\mathcal Q$. By Lemma~\ref{lem:row-support}(2), every inclusion-minimal initial node among the supporting rows must be a negative coordinate of $\widetilde r$, hence must be either $\mathcal A$ or $\mathcal J$. Because $\mathcal A\subsetneq\mathcal J$, these two sets cannot both be minimal. There is therefore a unique minimal initial node. If it were $\mathcal J$, every supporting initial node would contain $\mathcal J$: in a finite partially ordered set, descending from any element reaches a minimal element. Every coordinate appearing in the supporting rows would then contain $\mathcal J$. This is impossible, since the target has coefficient $+1$ at $\mathcal P,\mathcal Q\subsetneq\mathcal J$.

The unique minimal initial node is thus $\mathcal A$, and every supporting initial node contains $\mathcal A$. Comparing the $\mathcal A$ column in~\eqref{eq:irredundancy-combination} gives
\begin{equation}\label{eq:bottom-weight}
\sum_{\substack{s\in\mathscr S\\\beta(s)=\mathcal A}}\lambda_s=1.
\end{equation}
In particular, there is a supporting row with initial node $\mathcal A$.

The target vanishes at both feasible test vectors $v^{\mathcal P}$ and $v^{\mathcal Q}$. Evaluating~\eqref{eq:irredundancy-combination} at either vector gives a sum of nonnegative terms with positive weights. Consequently, every supporting row satisfies
\[
s(v^{\mathcal P})=s(v^{\mathcal Q})=0.
\]
Take a supporting row with initial node $\mathcal A$. As $|\mathcal E(\mathcal A)|\ge2$, it must be a four-term row, with a distinct cover pair $\{\mathcal U,\mathcal V\}$. By~\eqref{eq:principal-slack}, vanishing at $v^{\mathcal P}$ forces $\mathcal U\subseteq\mathcal P$ or $\mathcal V\subseteq\mathcal P$. Distinct covers are incomparable, so the pair must contain $\mathcal P$. Vanishing at $v^{\mathcal Q}$ likewise forces it to contain $\mathcal Q$. Thus $\{\mathcal U,\mathcal V\}=\{\mathcal P,\mathcal Q\}$ and $s=r$, contradicting the exclusion of $r$ from $\mathscr S$.

Now suppose that $r(h)=h_{\mathcal P}-h_{\mathcal A}$ with $\mathcal E(\mathcal A)=\{\mathcal P\}$. Its only negative complete coefficient is at $\mathcal A$. Lemma~\ref{lem:row-support}(2) therefore makes $\mathcal A$ the unique minimal supporting initial node. All supporting initial nodes contain it, and comparison at its column again yields~\eqref{eq:bottom-weight}. However, System~\ref{const:final} has only one row with this initial node, namely $r$ itself. This is the same contradiction.

Both arguments include $\mathcal A=\emptyset$, since the empty-set column was restored by Lemma~\ref{lem:row-support}(1). They also include $n=2$, when the system consists of one two-term inequality.
\end{proof}

\begin{corollary}[Full dimension and facets]\label{cor:final-facets}
For every $n\ge2$, $\mathscr C_3$ has dimension $D=|\mathcal{CS}|-1$ and satisfies no nonzero linear equality in these coordinates. Each distinct inequality in System~\ref{const:final} defines a facet of $\mathscr C_3$, that is, a face of dimension $D-1$.
\end{corollary}
\begin{proof}
Fix $0<\varepsilon<1/2$ and set
\begin{equation}\label{eq:strict-point}
h^*_{\mathcal C}=1-\varepsilon^{|\mathcal C|}\qquad(\mathcal C\in\mathcal{CS}).
\end{equation}
Then $h^*_\emptyset=0$. For a four-term row, $|\mathcal P|,|\mathcal Q|\ge|\mathcal A|+1$ gives
\[
\begin{aligned}
r(h^*)&=\varepsilon^{|\mathcal A|}-\varepsilon^{|\mathcal P|}
-\varepsilon^{|\mathcal Q|}+\varepsilon^{|\mathcal P\vee\mathcal Q|}\\
&\ge\varepsilon^{|\mathcal A|}(1-2\varepsilon)
+\varepsilon^{|\mathcal P\vee\mathcal Q|}>0.
\end{aligned}
\]
For a two-term row, $r(h^*)=\varepsilon^{|\mathcal A|}-\varepsilon^{|\mathcal P|}>0$. Since the system is finite, an open ball around $h^*$ still satisfies every inequality strictly. This proves full dimensionality and excludes all nonzero implicit linear equalities, including implicit equalities in the sense of Definition~\ref{def:implicit-redundant}.

For a fixed row $r$, Theorem~\ref{thm:general-irredundancy} supplies a vector $z$ with $s(z)\ge0$ for every $s\ne r$ and $r(z)<0$. Let
\[
t=\frac{r(h^*)}{r(h^*)-r(z)}\in(0,1),\qquad y=(1-t)h^*+tz.
\]
Then $r(y)=0$, while $s(y)=(1-t)s(h^*)+ts(z)>0$ for every $s\ne r$. A relative open neighborhood of $y$ in the hyperplane $r(h)=0$ therefore remains feasible. The face on this hyperplane has dimension $D-1$, proving the facet assertion.
\end{proof}

Together, Corollary~\ref{cor:equiv-final}, Theorem~\ref{thm:general-irredundancy}, and Corollary~\ref{cor:final-facets} show that System~\ref{const:final} is an equivalent, irredundant facet representation of System~\ref{const:initial} in the nonempty closed-set coordinates. This concerns the repair-constrained Shannon-type cone specified in Section~\ref{sec:mainidea}; adding reconstruction, symmetry, or other constraints may create further equalities or redundancies. Irredundancy here concerns the halfspace representation in these coordinates; it does not require linearly independent coefficient rows and makes no comparison with formulations using additional variables.

Table~\ref{tab:comparison} records the coordinate and constraint counts for $n\in\{2,3,4,5\}$. Irredundancy and the absence of implicit equalities follow for all $n\ge2$ from Theorem~\ref{thm:general-irredundancy} and Corollary~\ref{cor:final-facets}; the table provides numerical checks of the system sizes. The coordinate counts below exclude $H_\emptyset=0$. The System~\ref{thm:eliminatedsys} counts are obtained by enumerating the single-index closure targets at each closed set, simplifying comparable cases to two-term inequalities, and counting identical coefficient expressions once. As an independent counting check, the closed sets and their cover targets were enumerated for $n=2,3,4,5$, yielding respectively $1$, $89$, $40{,}391$, and $39{,}000{,}979$ System~\ref{const:final} inequalities, in agreement with the table.

\begin{table}[H]

		\caption{Coordinate and inequality counts; the fixed coordinate $H_\emptyset=0$ is excluded}
		\label{tab:comparison}
		\footnotesize
		\setlength{\tabcolsep}{3pt}
		\centering
		\begin{tabular}{|c|c|c|c|c|c|}
			\hline
			\multirow{2}{*}{$n$}  &  \multicolumn{2}{c|}{Coordinates} &  \multicolumn{3}{c|}{Inequalities}   \\
			\cline{2-6}
			& System~\ref{const:initial} & System~\ref{const:final} & System~\ref{const:initial} & System~\ref{thm:eliminatedsys} & System~\ref{const:final} \\
			\hline
			2 & 15 & 1 & 28 & 1 & 1  \\ \hline
			3 & 511 & 39 & 4,617 & 329 & 89 \\ \hline
			4 & 65,535 & 3,361 & 1,966,096 & 118,529 & 40,391\\ \hline
			5 & 33,554,431 & 969,375 & 2,516,582,425 & 93,171,259 &39,000,979 \\ \hline
		\end{tabular}

	\end{table}

\section{Construction of Inseparable Edges}\label{sec:edge}

\emph{System~\ref{const:final}} is defined by closed sets and their inseparable edges. We now represent the missing entries of a fixed closed set by a second graph. {This representation describes closure propagation and identifies the isolated directed cycles that generate nonsingleton covers. The same graph will be used for the counting bound in Section~\ref{sec:countbound}.}

\begin{define}[Off-diagonal complement digraph]\label{def:complement-digraph}
For a closed set $\mathcal A$, its \emph{off-diagonal complement digraph} $\Gamma_{\mathcal A}$ has vertex set $[n]$ and an arc $p\to q$ precisely when $p\ne q$ and the index $\iota(p,q):=n(p-1)+q$ is absent from $\mathcal A$. Let $R(\mathcal A)$ be the indices of the nonfull rows, and write $a_p$ and $b_p$ for the outdegree and indegree of vertex $p$ in this graph. An \emph{isolated directed cycle} is a directed cycle of length at least two whose vertices all have global indegree and outdegree one in $\Gamma_{\mathcal A}$.
\end{define}

The vertices of $\Gamma_{\mathcal A}$ are row indices, and its arcs identify missing off-diagonal variables for one fixed $\mathcal A$. By comparison, each vertex of the closed-set graph in Fig.~\ref{fig:graphfull} is an entire closed set, and each edge represents a single-index extension followed by closure. The latter graph is acyclic, since sets strictly increase along its edges.

For a closed $\mathcal A$, row $p$ is full exactly when its diagonal index is present. A nonfull row can have $a_p=0$ if only its diagonal is missing; however, it must have $b_p\ge1$, since a complete column would force that row to be full. Completing row $p$ removes all arcs leaving $p$. If this removes the last missing entry in a column $q$ whose row is not full, the column criterion completes row $q$. Thus propagation follows the direction of the arcs of $\Gamma_{\mathcal A}$.

For any index set $\mathcal U$, repeated applications of the row and column criteria give a \emph{closure chain}
\[
\mathcal U=\mathcal U_0\xrightarrow{\mathcal S_1}\mathcal U_1
\xrightarrow{\mathcal S_2}\cdots\xrightarrow{\mathcal S_m}\mathcal U_m
=\overline{\mathcal U},
\qquad \mathcal S_u=\mathcal U_u\setminus\mathcal U_{u-1}.
\]
Row criteria need only be applied separately to diagonals present at the start: each column step already completes its entire corresponding row, so its new diagonal does not require a further row step. For the construction of edges, we use a closed base $\mathcal A$ and a seed $i\notin\mathcal A$, and denote the chain starting at $\mathcal A\cup\{i\}$ by $\chain{i,\mathcal A}$.

\begin{define}[Active indices and active chains]\label{def:active-chain}
Fix a closed base $\mathcal A$. A missing index $x\notin\mathcal A$ is \emph{active relative to $\mathcal A$} if $\mathcal A\cup\{x\}$ is not closed; otherwise it is \emph{non-active}. Equivalently, for $x=\iota(p,q)$,
\[
x\text{ is active relative to }\mathcal A
\quad\Longleftrightarrow\quad
\begin{cases}
a_p>0,&p=q,\\
q\in R(\mathcal A)\text{ and }b_q=1,&p\ne q.
\end{cases}
\]
For $\mathcal B=\overline{\mathcal A\cup\{i\}}$, the chain $\chain{i,\mathcal A}$ is \emph{active} if every index in $\mathcal B\setminus\mathcal A$, including the seed, is active relative to $\mathcal A$.
\end{define}

The degree criteria follow directly from the repair rules. Adding a missing diagonal propagates exactly when its row still lacks an off-diagonal entry. Adding a missing off-diagonal entry propagates exactly when it completes a column whose row is not full. Activity is always evaluated against the original closed base, rather than against an intermediate set or a particular choice of the next step. In particular, every non-active index gives a singleton closed extension.

Example~\ref{ex:constructionofclosure} can be viewed as adding seed $1$ to the closed base $\{7,15\}$ for $n=4$. Among the subsequently added indices, $3$ and $11$ are active relative to that base, whereas $2,4,9,10,12$ are non-active. The chain is therefore not active. This conclusion is independent of the order in which eligible column criteria are applied.

For an isolated cycle $C:r_1\to\cdots\to r_\ell\to r_1$ in $\Gamma_{\mathcal A}$, write $r_{\ell+1}=r_1$ and let
\[
\Delta(C)=\{\iota(r_t,r_t),\ \iota(r_t,r_{t+1}):1\le t\le\ell\}
\]
be the set of its missing diagonal and off-diagonal indices.

\begin{proposition}[Active chains and isolated cycles]\label{prop:active-cycles}
For a closed $\mathcal A$ and $i\notin\mathcal A$, the chain $\chain{i,\mathcal A}$ is active if and only if $i\in\Delta(C)$ for an isolated directed cycle $C$ of $\Gamma_{\mathcal A}$. In that case,
\[
\overline{\mathcal A\cup\{i\}}=\mathcal A\cup\Delta(C).
\]
\end{proposition}
\begin{proof}
Suppose the chain is active, and set $\mathcal B=\overline{\mathcal A\cup\{i\}}$. Let $K$ be the indices of the newly completed rows, equivalently the rows whose diagonals belong to $\mathcal B\setminus\mathcal A$. This set is nonempty because the seed is active and every nontrivial repair step completes a row. For each $p\in K$, its diagonal is active relative to $\mathcal A$, so $a_p\ge1$.

All originally missing off-diagonal entries of row $p$ occur in $\mathcal B\setminus\mathcal A$, since $\mathcal B$ is closed and contains its diagonal. Each such entry $(p,q)$ is active, so $q\in R(\mathcal A)$ and $b_q=1$. Its inclusion completes column $q$ in $\mathcal B$, and hence $q\in K$. Consequently, every arc leaving $K$ ends in $K$, and each vertex in the induced subgraph on $K$ has outdegree at least one and indegree at most one. Counting these internal arcs gives
\[
|K|\le\sum_{p\in K}a_p\le |K|.
\]
Thus every vertex receives one internal arc, whose head was shown to have global indegree one, and every vertex has global outdegree one. The induced graph consists of disjoint isolated directed cycles.

A single seed can initiate only one of these isolated components. For a diagonal seed, its row belongs to $K$. For an off-diagonal seed, its head belongs to $K$; its tail must also belong to $K$, because each vertex of $K$ already receives its unique incoming arc from $K$. Propagation cannot leave that component, so $K$ is one cycle $C$ and $\mathcal B\setminus\mathcal A=\Delta(C)$.

Conversely, take an isolated cycle $C$ and any $i\in\Delta(C)$. Adding a diagonal completes its row and its unique missing outgoing entry. Adding an off-diagonal entry completes its head column and then its head row. In either case, propagation follows $C$ and fills all of $\Delta(C)$. No outside row is affected, because the cycle has no external incident arcs. The terminal set is $\mathcal A\cup\Delta(C)$, and every index of $\Delta(C)$ is active relative to $\mathcal A$. Hence the chain is active.
\end{proof}

\begin{theorem}[Inseparable edges]\label{thm:inseparableedge}
Given a closed set $\Acal$ and an element $i\in\Ncal\setminus\Acal$, the edge $\Acal\rightarrow\overline{\{i\}\cup\Acal}$ is inseparable if and only if either $\{i\}\cup\Acal$ is a closed set, or the closure chain $\chain{i,\Acal}$ is active.
\end{theorem}
\begin{proof}
Let $\mathcal B=\overline{\mathcal A\cup\{i\}}$. If the edge is inseparable and $\mathcal B\ne\mathcal A\cup\{i\}$, then $|\mathcal B\setminus\mathcal A|\ge2$. No $x\in\mathcal B\setminus\mathcal A$ can be non-active: otherwise $\mathcal A\cup\{x\}$ would be a proper intermediate closed set. Thus all indices of the difference are active, and so is the chain.

Conversely, a singleton closed extension is a cover. {If the chain is active, Proposition~\ref{prop:active-cycles} gives an isolated cycle $C$ with $\mathcal B=\mathcal A\cup\Delta(C)$. The same proposition applies to every $x\in\Delta(C)$, giving $\overline{\mathcal A\cup\{x\}}=\mathcal B$.} Any intermediate closed set containing such an $x$ would therefore contain $\mathcal B$, which is impossible. Hence the edge is inseparable.
\end{proof}

The closure formula in Proposition~\ref{prop:active-cycles} explains why the different seeds associated with one isolated cycle give a single cover target. Under System~\ref{const:initial}, Lemma~\ref{thm:closure} also gives
\[
H_{\mathcal A\cup\{x\}}=H_{\mathcal A\cup\Delta(C)}
\qquad(x\in\Delta(C)),
\]
so these seeds correspond to one closed entropy coordinate in Systems~\ref{thm:eliminatedsys} and~\ref{const:final}.

\begin{example}[{Seed propagation on an isolated cycle}]\label{ex:complement-digraph}
{Let $n=3$ and $\mathcal A=\{2,6,7\}$. Its absent off-diagonal indices are $3,4,8$, so $\Gamma_{\mathcal A}$ is the isolated cycle $1\to3\to2\to1$ in Fig.~\ref{fig:complement-digraph}, with $\Delta(C)=\{1,3,4,5,8,9\}$. A diagonal seed $1$ completes rows $1,3,2$ in that order. An off-diagonal seed $3$ first completes column $3$ and then rows $3,2,1$. In both cases, Proposition~\ref{prop:active-cycles} gives the same target $\mathcal A\cup\Delta(C)=\mathcal N$; the starting seed changes the propagation order but not this closed extension.}

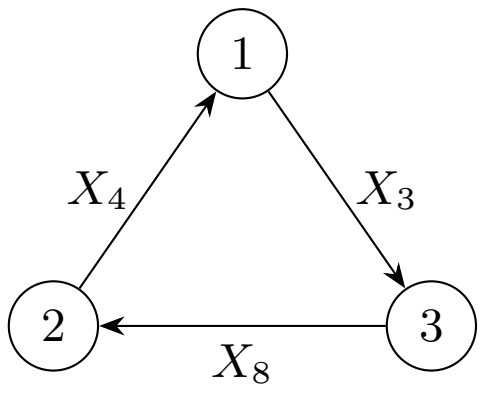
\begin{figure}[H]
\centering
\begin{minipage}[c]{.45\linewidth}
\centering
\[
\begin{pmatrix}
1&\boxed{2}&3\\4&5&\boxed{6}\\\boxed{7}&8&9
\end{pmatrix}
\]
\small (a) Boxed indices form $\mathcal A$.
\end{minipage}\hfill
\begin{minipage}[c]{.45\linewidth}
\centering
\begin{tikzpicture}[>=Stealth,every node/.style={font=\small}]
\node[draw,circle] (r1) at (0,1.1) {$1$};
\node[draw,circle] (r3) at (1.25,-.7) {$3$};
\node[draw,circle] (r2) at (-1.25,-.7) {$2$};
\draw[->] (r1)--node[right] {$X_3$}(r3);
\draw[->] (r3)--node[below] {$X_8$}(r2);
\draw[->] (r2)--node[left] {$X_4$}(r1);
\end{tikzpicture}
\par\small (b) $\Gamma_{\mathcal A}$ on the row indices.
\end{minipage}
\caption{The off-diagonal complement digraph for $n=3$ and $\mathcal A=\{2,6,7\}$. Arc labels identify the absent off-diagonal variables.}
\label{fig:complement-digraph}
\end{figure}

For comparison, the closed set $\mathcal A'=\{6,7,8\}$ and seed $2$ give the added sets $\{4,5\}$, $\{1,3\}$, and $\{9\}$. Here the original third row has no missing off-diagonal entry, so $9$ is non-active and $\mathcal A'\cup\{9\}$ is closed. It is a proper intermediate closed set below $\overline{\mathcal A'\cup\{2\}}=\mathcal N$, so this extension is separable. A newly added diagonal is therefore not automatically active.
\end{example}

Let $(\mathrm{bool},\mathrm{reason})=\textsf{IsInseparable}(i,\mathcal A)$ denote the edge test. A non-active seed gives a singleton cover. An active seed gives an inseparable edge exactly when its diagonal position or off-diagonal arc belongs to an isolated directed cycle of $\Gamma_{\mathcal A}$, by Proposition~\ref{prop:active-cycles} and Theorem~\ref{thm:inseparableedge}. When the edge is separable, $\mathrm{reason}$ may collect any non-active indices discovered during the closure construction. Each such index $b$ satisfies that $\mathcal A\cup\{b\}$ is closed and therefore gives a singleton cover. Returning an empty set of reasons is also valid; it only forgoes this shortcut.

Algorithm~\ref{algo:inseparableedges} enumerates all inseparable targets using this test. Every skipped index in $\mathrm{reason}$ has already contributed its target. Algorithm~\ref{algo:generatesystem} then generates the constraints of the directly defined System~\ref{const:final}. These are structural generation procedures, rather than iterative redundancy tests. The top set is excluded from the queue, so every processed closed set has at least one cover and the unique-target branch is well defined.

\begin{algorithm}[!htbp]
	\caption{\textsf{InseparableEdges}}
	\label{algo:inseparableedges}
	\KwIn{A closed set $\mathcal{A}$}
	\KwOut{The set $res$ containing all target nodes of inseparable edges originating from $\mathcal{A}$}

	Initialize $res \gets \emptyset$ and $\mathcal{S} \gets \mathcal{N} \setminus \mathcal{A}$\;

	\While{$\mathcal{S} \neq \emptyset$}{
		Select the minimal element $a \in \mathcal{S}$\;
		$\mathcal{S} \gets \mathcal{S} \setminus \{a\}$\;

		$(\mathrm{bool}, \mathrm{reason}) \gets \textsf{IsInseparable}(a, \mathcal{A})$\;

		\eIf{$\mathrm{bool}$}{
			$res \gets res \cup \{\overline{\{a\} \cup \mathcal{A}}\}$\;
		}{
			\ForEach{$b \in \mathrm{reason}$}{
				$res \gets res \cup \{\overline{\{b\} \cup \mathcal{A}}\}$\;
			}
			$\mathcal{S} \gets \mathcal{S} \setminus \mathrm{reason}$\;
		}
	}
	\Return{$res$\;}
\end{algorithm}

\begin{algorithm}[!htbp]
	\caption{\textsf{GenerateFinalSystem}}
	\label{algo:generatesystem}
	\KwIn{The number $n\ge2$ of storage nodes}
	\KwOut{The set $res$ of linear inequalities defining \emph{System~\ref{const:final}}}

	Initialize $\mathcal{N} \gets \{1, 2, \ldots, n^2\}$\;
	Initialize $\textsf{org\_node}\gets\{\emptyset\}$ and $res\gets\emptyset$\;
Initialize $\textsf{seen}\gets\{\emptyset\}$\;

	\While{$\textsf{org\_node} \neq \emptyset$}{
		Select a node $\mathcal{A} \in \textsf{org\_node}$\;
		$\textsf{org\_node} \gets \textsf{org\_node} \setminus \{\mathcal{A}\}$\;

		$\textsf{end\_node} \gets \textsf{InseparableEdges}(\mathcal{A})$\;

		\eIf{$|\textsf{end\_node}| > 1$}{
			\ForEach{unordered pair $(\mathcal{B}, \mathcal{C})$ with $\mathcal{B}, \mathcal{C} \in \textsf{end\_node}$ and $\mathcal{B} \neq \mathcal{C}$}{
				Append $H_{\mathcal{B}} + H_{\mathcal{C}} - H_{\overline{\mathcal{B} \cup \mathcal{C}}} - H_{\mathcal{A}} \ge 0$ to $res$\;
			}
		}{
			Let $\mathcal{B}$ be the unique node in $\textsf{end\_node}$\;
			Append $H_{\mathcal{B}} - H_{\mathcal{A}} \ge 0$ to $res$\;
		}

		$\textsf{new}\gets\textsf{end\_node}\setminus(\textsf{seen}\cup\{\mathcal N\})$\;
$\textsf{org\_node}\gets\textsf{org\_node}\cup\textsf{new}$\;
$\textsf{seen}\gets\textsf{seen}\cup\textsf{new}$\;
	}
	\Return{$res$\;}
\end{algorithm}

\section{A Computable Bound from Local Degrees}
\label{sec:countbound}
For a closed set $\mathcal A$, write $d(\mathcal A)$ for the number of outgoing inseparable edges. The construction of \emph{System~\ref{const:final}} generates
\begin{equation}\label{eq:qdegree}
q(d)=\begin{cases}
0,&d=0,\\
1,&d=1,\\
\binom d2,&d\ge2
\end{cases}
\end{equation}
constraints at $\mathcal A$. In particular, a single outgoing edge still produces one two-term inequality. The cardinality bound
\[
U_n=\sum_{\mathcal A\in\CScal}\binom{n^2-|\mathcal A|}{2}
\]
treats every missing element as a possible distinct cover. A much tighter estimate follows by separating singleton extensions from the nontrivial cycles in the complement of a closed set.

\Needspace{12\baselineskip}

We use the off-diagonal complement digraph $\Gamma_{\mathcal A}$ of Definition~\ref{def:complement-digraph}, with the degrees $a_i,b_i$ measured before any seed is added, and write $R=R(\mathcal A)$. Let $c(\mathcal A)$ count its isolated directed cycles. Define

\begin{align*}
&M_{\rm off}=|E(\Gamma_{\mathcal A})|,
&&{\text{number of missing off-diagonal entries}},\\
&c_1=|\{i\in R:b_i=1\}|,
&&{\text{number of columns $i\in R$ with one missing off-diagonal entry}},\\
&p=|\{i\in R:a_i=0\}|,
&&{\text{number of nonfull rows missing only their diagonal}},\\
&s(\mathcal A)=M_{\rm off}-c_1+p,
&&{\text{number of singleton closed extensions}},\\
&v(\mathcal A)=|\{i\in R:a_i=b_i=1\}|,
&&{\text{number of candidate vertices for isolated cycles}}.
\end{align*}
{The columns counted by $c_1$ are indexed only by nonfull rows: completing a column indexed by an already full row cannot cause further propagation. The quantities $M_{\rm off}-c_1$ and $p$ count the non-active off-diagonal and diagonal indices, respectively. The vertices counted by $v$ need not all lie on isolated cycles; they provide the candidate vertices used in the bound below.}

Theorem~\ref{thm:inseparableedge} and Proposition~\ref{prop:active-cycles} distinguish the singleton covers, generated by non-active indices, from the nonsingleton covers, generated by isolated cycles. In Example~\ref{ex:complement-digraph},
\[
R=\{1,2,3\},\quad M_{\rm off}=3,\quad c_1=3,\quad p=0,
\qquad s(\mathcal A)=0,\quad v(\mathcal A)=3,\quad c(\mathcal A)=1.
\]
Thus there is one nonsingleton cover and no singleton cover: $\mathcal E(\mathcal A)=\{\mathcal N\}$. The local count below gives $d(\mathcal A)=1$, and the corresponding System~\ref{const:final} constraint is $H_{\mathcal N}-H_{\mathcal A}\ge0$.

\Needspace{4\baselineskip}
For the closed set $\mathcal A'=\{6,7,8\}$ in the same example, $s(\mathcal A')=1$ and $c(\mathcal A')=0$; its unique cover is $\mathcal A'\cup\{9\}$. The same degree $d(\mathcal A')=1$ is therefore contributed by a singleton extension.

\begin{lemma}[Local cover count]\label{lem:local-cover-count}
The number $d(\mathcal A)$ of outgoing inseparable edges is
\begin{equation}\label{eq:local-cover-count}
d(\mathcal A)=s(\mathcal A)+c(\mathcal A).
\end{equation}
Consequently,
\begin{equation}\label{eq:local-cover-upper}
d(\mathcal A)\le b(\mathcal A):=s(\mathcal A)+\left\lfloor\frac{v(\mathcal A)}2\right\rfloor.
\end{equation}
\end{lemma}

\begin{proof}
By Definition~\ref{def:active-chain}, a missing diagonal gives a singleton cover exactly when its row has no missing off-diagonal entry, contributing $p$. A missing off-diagonal entry gives a singleton cover unless its head belongs to a nonfull row and has indegree one. Each such critical head excludes exactly one entry, so the off-diagonal singleton covers number $M_{\rm off}-c_1$. This proves that $s(\mathcal A)$ counts the singleton covers.

Every nonsingleton cover is generated by an active chain, by Theorem~\ref{thm:inseparableedge}, and hence by an isolated directed cycle, by Proposition~\ref{prop:active-cycles}. Conversely, each isolated cycle generates one such cover. {All seeds in the index set $\Delta(C)$ of one isolated cycle give the same target by Proposition~\ref{prop:active-cycles}, whereas different isolated cycles give different targets.} Thus there are exactly $c(\mathcal A)$ nonsingleton covers, proving~\eqref{eq:local-cover-count}.

The isolated cycles are vertex-disjoint, and each uses at least two vertices with $a_i=b_i=1$. Therefore $c(\mathcal A)\le\lfloor v(\mathcal A)/2\rfloor$, proving~\eqref{eq:local-cover-upper}.
\end{proof}

The exact local count in Lemma~\ref{lem:local-cover-count} gives
\begin{equation}\label{eq:exact-final-count}
N_{\rm Final}=\sum_{\mathcal A\in\mathcal{CS}}q\bigl(s(\mathcal A)+c(\mathcal A)\bigr).
\end{equation}
There is no duplication in this sum. For a four-term inequality, its two positive coordinates recover the unordered pair of cover targets. The negative coordinates identify the initial node and the join, since the former is a strict subset of the latter; an absent initial-node coordinate means $\mathcal A=\emptyset$. A two-term inequality is determined by its initial node and unique cover target and cannot coincide with a four-term one. By Theorem~\ref{thm:general-irredundancy} and Corollary~\ref{cor:final-facets},~\eqref{eq:exact-final-count} is also the exact number of facet inequalities in this representation.

To compute $c(\mathcal A)$, consider the subgraph of $\Gamma_{\mathcal A}$ induced by the vertices with $a_i=b_i=1$. Its directed cycles are precisely the isolated cycles of Definition~\ref{def:complement-digraph}, so a traversal counts them without generating closure targets. The degree bound below avoids this traversal by replacing $c(\mathcal A)$ with $\lfloor v(\mathcal A)/2\rfloor$.

Since $q$ in~\eqref{eq:qdegree} is nondecreasing, a sharper computable bound is
\begin{equation}\label{eq:tight-count-bound}
N_{\rm Final}\le T_n:=\sum_{\mathcal A\in\CScal}q\!\left(s(\mathcal A)+\left\lfloor\frac{v(\mathcal A)}2\right\rfloor\right).
\end{equation}
The summands require only the row and column degrees of $\Gamma_{\mathcal A}$, without closure generation or edge enumeration. 
To make the domains in this sum explicit, first choose the $k$ full rows of a proper closed set. As in Theorem~\ref{thm:numofclosures}, $0\le k\le n-2$; put $r=n-k$. A simultaneous permutation of the row and column labels preserves all the quantities in the summand. We may therefore fix the nonfull labels to be $R=[r]$ and the full labels to be $F=\{r+1,\ldots,n\}$, and multiply by $\binom nk$ for the choices of the full rows. A column indexed by $F$ need not itself be complete; its corresponding row is already full.

The remaining choices specify the \emph{missing off-diagonal entries} in the nonfull rows. They split into two blocks:
\begin{enumerate}[(1)]
\item In the $r\times k$ block with row label $i\in R$ and column label $r+\ell\in F$, choose any subset
\[
X\subseteq[r]\times[k].
\]
The pair $(i,\ell)\in X$ means that the original array position $(i,r+\ell)$, with index $\iota(i,r+\ell)$, is absent. Thus the second coordinate $\ell$ is a local label for a column indexed by $F$. There is no restriction on $X$, since completing a column in $F$ cannot add a row that is not already full. For $k=0$, this block is empty and $X=\emptyset$ is its only choice.
\item For each nonfull column $j\in R$, choose independently a nonempty set
\[
\emptyset\ne Y_j\subseteq[r]\setminus\{j\}.
\]
Here $Y_j$ is a set of \emph{row labels}: $i\in Y_j$ means that position $(i,j)$, with index $\iota(i,j)$, is absent. The label $j$ is excluded because these are off-diagonal positions. Nonemptiness prevents the column criterion from filling row $j$, which is required to remain nonfull.
\end{enumerate}
All entries in the full rows are present, and all $r$ diagonals in the nonfull rows are absent. Consequently, these choices determine exactly one closed set:
\[
\mathcal A(X,Y)=\mathcal N\setminus\left(
\{\iota(i,i):i\in[r]\}
\cup\{\iota(i,r+\ell):(i,\ell)\in X\}
\cup\bigcup_{j=1}^{r}\{\iota(i,j):i\in Y_j\}
\right),
\]
where $Y=(Y_1,\ldots,Y_r)$ is the ordered tuple of column choices. Conversely, a closed set with these full-row labels uniquely specifies $X$ and $Y$. Thus there are $2^{rk}$ choices for $X$ and $(2^{r-1}-1)^r$ choices for $Y$, recovering the corresponding factors in Theorem~\ref{thm:numofclosures}.

The degree data used in the summand can be read directly from these sets. For $i,j\in[r]$,
\[
\begin{aligned}
a_i&=|\{\ell\in[k]:(i,\ell)\in X\}|+|\{j\in[r]:i\in Y_j\}|,\\
b_j&=|Y_j|,\qquad M_{\rm off}=|X|+\sum_{j=1}^{r}|Y_j|.
\end{aligned}
\]
In particular, $c_1=|\{j\in[r]:|Y_j|=1\}|$; the definitions of $p$, $s$, and $v$ then give $s(X,Y)$ and $v(X,Y)$. Substituting into~\eqref{eq:tight-count-bound} yields
\begin{equation}\label{eq:tight-count-finite-sum}
T_n=\sum_{k=0}^{n-2}\binom nk
\sum_{X\subseteq[r]\times[k]}\quad
\sum_{\substack{\emptyset\ne Y_j\subseteq[r]\setminus\{j\}\\j=1,\ldots,r}}
q\!\left(s(X,Y)+\left\lfloor\frac{v(X,Y)}2\right\rfloor\right),
\qquad r=n-k.
\end{equation}
The middle sum runs over every missing-entry pattern $X$ in the cross-block. The innermost sum runs over every ordered tuple $(Y_1,\ldots,Y_r)$ satisfying the stated condition separately for each column; it is not a sum over a single column label $j$. The full closed set $\mathcal N$ contributes zero and is omitted.

For example, take $n=4$, $k=1$, $r=3$, with full row $4$, and choose
\[
X=\{(1,1)\},\qquad Y_1=\{2\},\quad Y_2=\{1,3\},\quad Y_3=\{1\}.
\]
The set $X$ removes position $(1,4)$; the three $Y_j$ remove positions $(2,1)$, $(1,2),(3,2)$, and $(1,3)$, respectively. Including the three absent diagonals, the resulting closed set is $\mathcal A=\{7,8,9,12,13,14,15,16\}$. Its nonfull-row outdegrees are $(a_1,a_2,a_3)=(3,1,1)$ and its corresponding indegrees are $(b_1,b_2,b_3)=(1,2,1)$. Hence $M_{\rm off}=5$, $c_1=2$, $p=0$, $s=3$, and $v=1$, giving the summand $q(3+\lfloor1/2\rfloor)=3$.

Using this explicit enumeration of closed sets, both the exact count~\eqref{eq:exact-final-count} and the degree bound~\eqref{eq:tight-count-bound} can be evaluated in $O(n^2|\mathcal{CS}|)$ arithmetic operations. For each closed set, scanning the missing-entry array constructs the degrees and the graph in $O(n^2)$ time; counting cycles in the induced graph above takes $O(n)$ further time because each vertex has at most one incoming and one outgoing arc. Thus the degree bound has the convenience of using only degrees, while both evaluations still sum over all closed sets. This is a counting cost, separate from the cost of writing every defining inequality.

\begin{proposition}[Comparison with the cardinality bound]\label{prop:bound-comparison}
For every $n\ge2$,
\begin{equation}\label{eq:bound-comparison}
N_{\rm Final}\le T_n\le U_n.
\end{equation}
\end{proposition}
\begin{proof}
The first inequality follows from Lemma~\ref{lem:local-cover-count} and the monotonicity of $q$. For the second, fix a proper closed set $\mathcal A$ and put $m=n^2-|\mathcal A|$. Each nonfull row lacks its diagonal, so
\[
m=M_{\rm off}+|R|.
\]
The vertices counted by $v$ are among those counted by $c_1$, and those counted by $p$ lie in $R$. Therefore
\[
\begin{aligned}
b(\mathcal A)&=M_{\rm off}-c_1+p+\lfloor v/2\rfloor\\
&\le M_{\rm off}+p\le M_{\rm off}+|R|=m.
\end{aligned}
\]
A proper closed set has at least two nonfull rows, by the argument in Theorem~\ref{thm:numofclosures}; hence $m\ge2$. It follows that $q(b(\mathcal A))\le q(m)=\binom m2$. The full closed set contributes zero to both bounds. Summing over all closed sets gives $T_n\le U_n$.
\end{proof}

\begin{table}[H]

\centering\small
\caption{Comparison of the cardinality and local-degree bounds.}
\label{tab:tight-count-bound}
\begin{tabular}{rrrrr}
\toprule
$n$ & $U_n$ & $T_n$ & System~\ref{const:final} count & Excess of $T_n$\\
\midrule
2 & 6 & 1 & 1 & $0\%$\\
3 & 699 & 95 & 89 & $6.74\%$\\
4 & 160,470 & 41,135 & 40,391 & $1.84\%$\\
5 & 104,464,390 & 39,185,434 & 39,000,979 & $0.47\%$\\
\bottomrule
\end{tabular}
\end{table}

The values in Table~\ref{tab:tight-count-bound} are obtained with exact integer arithmetic. For $n=4$ and $n=5$, $T_n$ exceeds the \emph{System~\ref{const:final}} count by only $744$ and $184,455$ constraints, respectively. An independent exhaustive calculation generates all distinct single-element closures, retains the minimal targets, and sums $q(d(\mathcal A))$; it agrees with~\eqref{eq:local-cover-count} for every closed set for $n=2,3,4,5$. This independently validates the reported counts.

\Needspace{12\baselineskip}
\section{Conclusion}
\label{sec:con}
We have described a closure-based representation of the Shannon-type constraint system for the $(n,n-1,n-1)$ exact-repair model. Nonempty closed sets index the reduced entropy coordinates, and inseparable edges specify \emph{System~\ref{const:final}} directly. For every $n\ge2$, the system is equivalent to \emph{System~\ref{const:initial}}, is full-dimensional in its nonempty closed-set coordinates, and has no redundant defining inequality or implicit equality. Each distinct inequality defines a facet of the repair-constrained Shannon-type cone. The proof uses the cover structure and feasible test vectors, independently of small-parameter computations.

The off-diagonal complement digraph gives an exact local cover count and a readily computed upper bound on the number of defining inequalities. At $n=4$ and $n=5$, this bound exceeds the actual counts by only $1.84\%$ and $0.47\%$, respectively. Exact enumeration checks the reported system sizes, while the general irredundancy theorem establishes an irredundant facet representation in the specified coordinate space. Implementation-level LP performance comparisons and the effect of additional coding constraints remain topics for further work.

\end{document}